\documentclass[aip,amsmath,amssymb,reprint]{revtex4-1}

\usepackage[utf8]{inputenc}
\usepackage{graphicx,color}
\usepackage{dcolumn}

\renewcommand{\Re}{\mbox{\it Re}}

\newcommand{\pcav}{p_{\textnormal{\scriptsize{cav}}}}
\newcommand{\pref}{p_{\textnormal{\scriptsize{ref}}}}

\begin{document}
\title[Lagrangian statistics of pressure fluctuations]{Lagrangian statistics of pressure fluctuation events in homogeneous isotropic turbulence}
\author{Mehedi Bappy}
\author{Pablo M. Carrica}
\affiliation{IIHR-Hydroscience and Engineering, The University of Iowa, Iowa City, IA, USA}
\author{Gustavo C. Buscaglia}
\affiliation{Instituto de Ci\^encias Matem\'aticas e de Computa\c{c}\~ao, Universidade de S\~ao Paulo, S\~ao Carlos, Brazil}

\begin{abstract}
Homogeneous and isotropic turbulent fields obtained from two Direct Numerical Simulation databases (with $\Re_\lambda$ equal to 150 and 418)  were seeded with point particles that moved with the local fluid velocity to obtain Lagrangian pressure histories. Motivated by cavitation inception modeling, the statistics of events in which such particles undergo low-pressure fluctuations were computed, parameterized by the amplitude of the fluctuations and by their duration. The main results are the average frequencies of these events and the probabilistic distribution of their duration, which are of predictive value. A connection is also established between these average frequencies and the pressure probability density function, thus justifying experimental methods proposed in the literature. Further analyses of the data show that the occurrence of very-low-pressure events is highly intermittent and is associated with worm-like vortical structures of length comparable to the integral scale of the flow.
\end{abstract}


\maketitle

\section{Introduction}

The study of pressure fluctuations in turbulent flows has been the subject of significant theoretical, numerical and experimental work over more than eight decades. Over the years, much knowledge has been gained about the intensity of the fluctuations, their scaling properties, their energy spectrum and probability density function (PDF) as well as about their spatial structure and their relation to other flow variables. The reader is referred, among others, to the works of George {\em et al} \cite{gba84}, Pumir \cite{pumir94_pf}, Cao {\em et al} \cite{ccd99_pf}, Gotoh \& Rogallo \cite{gr99_jfm} and references therein for further details.

Pressure fluctuations have important practical consequences in many physical situations, since they intervene in the forces exerted by the fluid on adjacent and immersed bodies and are also key to acoustic noise. The specific motivation of the study reported herein is the phenomenon of incipient cavitation in turbulent liquid flows, which is a macroscopic consequence of pressure fluctuations on microscopic bubbles (or {\em nuclei}) present in the liquid\cite{Brennen1995,morch15_if}. Though the dynamics of cavitation bubbles has been well established following the pioneering work of Plesset \cite{plesset49_jam}, accurate prediction of cavitation inception in technologically relevant flows remains a challenge\cite{lbm15_if,li15_if}.

A seminal work on cavitation inception in turbulent flows was published in 1979 by Arndt \& George \cite{ag79}. It puts forward the main ingredients for turbulence-induced inception: That the cavitation nuclei are subject to pressure fluctuations as measured in a {\em Lagrangian frame}, and that (a) the pressure must dip below a critical level ($p_{\mbox{\scriptsize{cav}}}$, the {\em cavitation pressure}, { also called {\em Blake threshold}}\cite{hbgl98_jfm}), and (b) the low pressure must persist below the critical level for a time that exceeds the time scale for bubble growth. The picture has been confirmed and enriched over the years by several authors \cite{rk94_jfm,bfm95_jfm,lvmb00_pf,ka02_prs}, including the role of coherent structures and the possibility of using incipient cavitation bubbles as pressure sensors.

Typically, cavitation inception is defined by the inception cavitation index $\sigma_i = (p_{\mbox{\scriptsize{ref}},i} - p_{\mbox{\scriptsize{cav}}})/(\frac12 \rho U^2)$, where $p_{\mbox{\scriptsize{ref}},i}$ is the value of the reference pressure of the flow at which cavitating nuclei first become {\em observable}, $\rho$ is the { liquid's} density and $U$ the global velocity scale. By observable it is meant that the {\em frequency} of cavitation events growing bubbles to a large enough size is high enough to be experimentally detectable. In view of criteria (a) and (b) above, inception modeling thus requires knowledge about {\em the frequency $\zeta$ at which random turbulent fluctuations take the pressure, as experienced by the minute nuclei transported by the flow, below $p_{\mbox{\scriptsize{cav}}}$ for long enough time}. Such knowledge is, however, not as yet available in the literature. The purpose of this contribution is to provide data about $\zeta$ in the simplest turbulent flow, homogeneous isotropic turbulence (HIT). { As such, the results are related to the so-called {\em homogeneous nucleation} of cavitation, in which the events take place far away from walls. To address {\em heterogeneous nucleation} one should increase at least one step further the complexity of the flow and consider boundary layer turbulence. 
  }

The pressure PDF in HIT is known to be negatively skewed and exhibit an exponential tail at very low pressures \cite{pumir94_pf,ccd99_pf,gr99_jfm}.
On the basis of physical intuition, several authors have argued that the frequency $\zeta(\delta p)$ of pressure fluctuations below $\delta p$ should also exhibit an exponential tail for large and negative $\delta p$ \cite{ag79,rk94_jfm,lvmb00_pf}. Mathematically, this is far from obvious. In fact, consider a $p(t)$ that exhibits logarithmic pulses of the form $\ln (t-t_i)$ where $t_1, t_2,\ldots$ are random times which arise with average frequency $f$. It is easy to check that in such a case $\mbox{PDF}(p)$ has an exponential tail for $p \ll 0$. However, the frequency $\zeta(\delta p)$ is not exponential, as it equals $f$ for any (sufficiently negative) $\delta p$. The temporal structure of the pressure excursions is thus crucial in determining the statistics of occurrence of low pressure events.

Having in mind cavitation modeling, in this investigation we perform Lagrangian sampling of two Direct Numerical Simulation (DNS) databases counting events at which the particles go below some given pressure threshold $\delta p$ for a longer time than some given minimum duration $d\geq 0$. Notice that each excursion, even if the pressure dives well below $\delta p$ for a time much greater than $d$, is counted as a single event. The rationale behind this is that, if the given $\delta p$ and $d$ are likely to produce cavitation of the particle (nucleus), then the gaseous phase will violently grow and increase the local pressure so that no similar event will take place within the same pressure excursion. The results obtained from { our} Lagrangian counting experiments confirm the exponential dependence $\zeta(\delta p) \simeq C \exp (\beta \delta p)$ for $\delta p \ll 0$. The factors $C$ and $\beta$ are retrieved by fitting the data. We also address the random structure of the pressure fluctuation events by building the PDF of interarrival times. It shows that that low-pressure events do not behave as a Poisson process, exhibiting a marked burstiness that increases as $\delta p$ becomes more negative. It is appropriate to point out that, since the tracking method does not incorporate any relative velocity between the Lagrangian particles and the fluid, the results only apply to the smallest cavitation nuclei.

\section{Definitions and methods}

Two databases containing Direct Numerical Simulation (DNS) results of Forced Homogeneous Isotropic Turbulence (FHIT) were queried. They consist of fully resolved numerical solutions of the incompressible Navier-Stokes equations in a periodic domain, with forcing applied to a narrow band of low wavenumber modes in such a way that a statistically steady flow develops.

The DNS results are time series of velocity and pressure spatial fields with zero mean from which statistical averages of kinetic energy $\langle K \rangle$ and viscous dissipation $\langle \epsilon \rangle$ were computed. From these values the basic scales characterizing each flow were defined as:
\begin{equation}
  \begin{cases}
    \mbox{Velocity scale: } & u' = \sqrt{\frac{2 \langle K \rangle}{3}}, \\
    \mbox{Length scale: } & \lambda = u' \sqrt{\frac{15 \nu}{\langle \epsilon \rangle}},
  \end{cases}
\end{equation}
where $\nu$ is the kinematic viscosity and unit density is used. The length scale so defined is the Taylor length microscale. With these two basic scales all variables were rescaled (non-dimensionalized), dividing velocities by $u'$, lengths by $\lambda$, pressures by { $\rho u'^2$}, times by $\lambda/u'$, etc. All reference to the DNS data made { throughout this article} concerns the scaled (dimensionless) variables. In particular, the scaled velocity field ${\bf u}$ satisfies
\begin{equation}
\langle \|{\bf u}\|^2 \rangle = 3, \qquad \langle \|\nabla \times {\bf u }\|^2 \rangle = 15. 
\end{equation}
The non-dimensional parameter that characterizes the flow is the Reynolds number $\Re_\lambda = u'\lambda / \nu$. Two values were considered: $\Re_\lambda = 150$, obtained from the database maintained by J. Jim\'enez and coworkers at Univ. Polit\'ecnica Madrid \cite{cvj17_science}, and $\Re_\lambda = 418$, obtained from the Johns Hopkins Turbulence Databases \cite{lietal08_jt,yuetal12_jt,fitdoi}.

Notice that, under this scaling, the Kolmogorov scales\cite{pope00_book} are given by
\begin{equation}
  \begin{cases}
    \mbox{Length:} & \eta_K=15^{-\frac14} \Re_\lambda^{-\frac12}=0.508\,\Re_\lambda^{-\frac12},\\
    \mbox{Velocity:} & u_K=15^{\frac14}\Re_\lambda^{-\frac12}=1.968 \Re_\lambda^{-\frac12},\\
    \mbox{Time:} & \tau_K={15^{-\frac12}}=0.2582.
  \end{cases}
  \end{equation}
  Notably, $\tau_K$ is independent of $\Re_\lambda$. The Reynolds number based on the Kolmogorov length is then $\Re_\eta= u'\eta_K/\nu= {15}^{-\frac14}\,\Re_\lambda^{\frac12}=0.508\Re_\lambda^{\frac12} $.

  The Lagrangian trajectories are computed by numerically solving the equation
  \begin{equation}
    \frac{d{\bf r}}{dt}(t)={\bf u}({\bf r}(t),t)
  \end{equation}
  from time $t=0$ to $t=T$, with initial condition ${\bf r}(0)={\bf X}$, for a large number $M$ of initial positions, i.e., ${\bf X}\in \mathcal{X}=\{ {\bf X}^1,{\bf X}^2,\ldots,{\bf X}^M\}$. The adopted time-integration method is a Runge-Kutta scheme of order $\geq 2$, with a time step much smaller than the Kolmogorov time scale ($\Delta t < 0.05 \tau_K$). The required spatial interpolation was accomplished using Lagrangian polynomial interpolants of degree three or greater, depending on the case.

  Along the set of computed trajectories several statistical estimates were computed by averaging over the Lagrangian samples (parameterized by the set $\mathcal{X}$ of initial positions) and over time. For these estimates to be meaningful recommended practices were followed { \cite{y02_arfm}}. The initial positions were uniformly distributed over the simulation domain, keeping them at least 2.5 grid spacings apart. The integration time $T$ extended over several eddy turnover times. Specifically, the $\Re_\lambda=150$ field (resolution 256$^3$,  time step 0.043) was sampled with $M=10^6$ particles during a time $T=869.1$ (114 turnover times). The $\Re_\lambda=418$ field (resolution 1024$^3$, time step 0.012), on the other hand, was sampled with $M=4\times 10^5$ particles during a time $T=60.7$, which corresponds to 5 turnover times.

  In incompressible turbulence Lagrangian averages of spatial quantities (e.g., kinetic energy, enstrophy, etc.) coincide with Eulerian averages. Due to the finite number of Lagrangian particles and the finite simulation time, especially in the case with $\Re_\lambda=418$, the estimates obtained by averaging over the particles and over time are not exact. The difference of our estimates with Eulerian averages can be used as a simple test of statistical significance. The error in the rms values of velocity and vorticity were smaller than 2\%, value recommended by Yeung \cite{y02_arfm}, providing some validation of our sampling procedure.

The Lagrangian ``Taylor'' timescale of a variable $Y$ is defined as \cite{y02_arfm}
\begin{equation}
  \tau_{Y}=\left [ \frac{\mbox{Var}(Y)}{\mbox{Var}(D Y/D t)} \right ]^{\frac12},
\end{equation}
where $\mbox{Var}(\cdot)$ denotes the variance of the corresponding quantity and $D/Dt$ is the material derivative. 
The Lagrangian autocorrelation function of a variable $Y$ with zero mean is defined by
\begin{equation}
\rho_Y(s)=\frac{\langle Y({\bf r}(t),t) Y({\bf r}(t+s),t+s)\rangle}{\mbox{Var}(Y)}.
\end{equation}
The Lagrangian integral timescale of $Y$, denoted by $\mathcal{T}(Y)$, is given by
\begin{equation}
  \mathcal{T}_Y=\int_0^{+\infty} \rho_Y(s)~ds~.
  \end{equation}
All the statistical averages in the previous and forthcoming definitions are of course replaced by estimates computed over the numerical approximation of the sampled trajectories.

Finally, the occurrence of low pressure events was investigated as follows: Some negative pressure values $p_-$ were chosen, namely -2, -2.05, -2.1, etc. Notice that these correspond to non-dimensional fluctuations, since $\langle p \rangle = 0$. For a given value of ${p_-}$, a Lagrangian particle was defined as undergoing a {\em low-pressure event with threshold $p_-$} that starts at time $t_{\mbox{\scriptsize{start}}}$ if its pressure satisfies ${p}(t_{\mbox{\scriptsize{start}}})={p_-}$ and $Dp/Dt(t_{\mbox{\scriptsize{start}}}) < 0$. The end of the event is defined as the first time $t_{\mbox{\scriptsize{end}}}$ such that $p(t_{\mbox{\scriptsize{end}}})=p_-$ and $Dp/Dt(t_{\mbox{\scriptsize{end}}})>0$.
The {\em duration} of the event is defined as $t_{\mbox{\scriptsize{end}}}-t_{\mbox{\scriptsize{start}}}$. 

In this way, given $m$ particles evolving in a turbulent field, a stochastic {\em counting process} $n(p_-,d,t)$ can be defined as the number of low-pressure events (with threshold $p_-$ and duration $t_{\mbox{\scriptsize{end}}}-t_{\mbox{\scriptsize{start}}}>d$) that have $0<t_{\mbox{\scriptsize{start}}}<t$.
For each threshold $p_-$ and each minimum duration $d$, the number $n(p_-,d,T)$ of events was computed from the time history of the $M$ Lagrangian particles that are traced over the time $T$ of the simulation. From this, we computed an estimate for the rate of such events as
  \begin{equation}
    \zeta(p_-,d)\simeq\frac{n(p_-,d,T)}{MT}~.
  \end{equation}
  An analogous processing was carried out for high-pressure events, denoting the (positive) threshold by $p_+$.

\section{Results}

\subsection{Basic statistics}

\begin{figure}[htp]
		\centering 
		\includegraphics[scale=0.35]{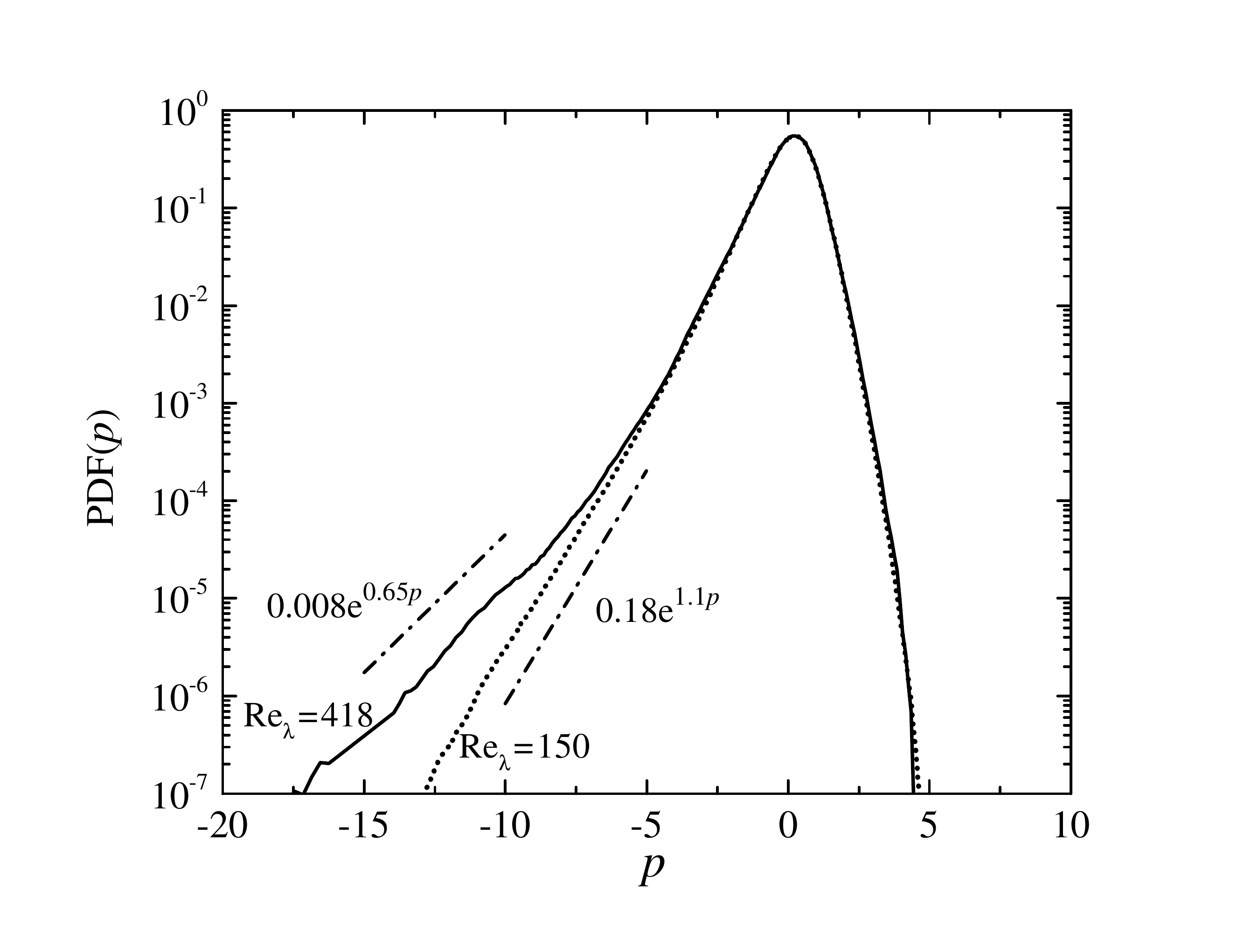}
		\caption{Pressure PDFs obtained from the Lagrangian data at $\Re_\lambda=150$ and $418$. { The dash-dotted lines depict the fitted exponentials of Eq. \ref{eqpdf}, shifted for clarity.}
                }\label{fig:pressurepdfs}
	\end{figure}

We first report some basic statistical quantities obtained from the computed Lagrangian pressure histories. These include the pressure PDFs shown in Fig. \ref{fig:pressurepdfs}, together with the quantities summarized Table \ref{tablebasic}.


\begin{table}
\begin{center}
\begin{tabular}{lccc}
  \hline
  Quantity & Symbol & $\Re_\lambda=150$ & $\Re_\lambda=418$ \\ \hline
  Variance of $p$ & Var$(p)$ & 0.78 & 0.82 \\
  Variance of $Dp/Dt$ & $~~~$Var$(Dp/Dt)$ & 1.48 & 0.36 \\
  Time microscale & $\tau_p$ & 0.73 & 1.51 \\
  Integral timescale & $\mathcal{T}_p$ & 2.30 & 6.21 \\ \hline
\end{tabular}
\end{center}
\caption{Basic statistics of the computed Lagrangian pressure histories.}\label{tablebasic}
\end{table}

\begin{figure}[htp]
		\centering 
		\def\svgwidth{\textwidth}	
		\includegraphics[scale=0.35]{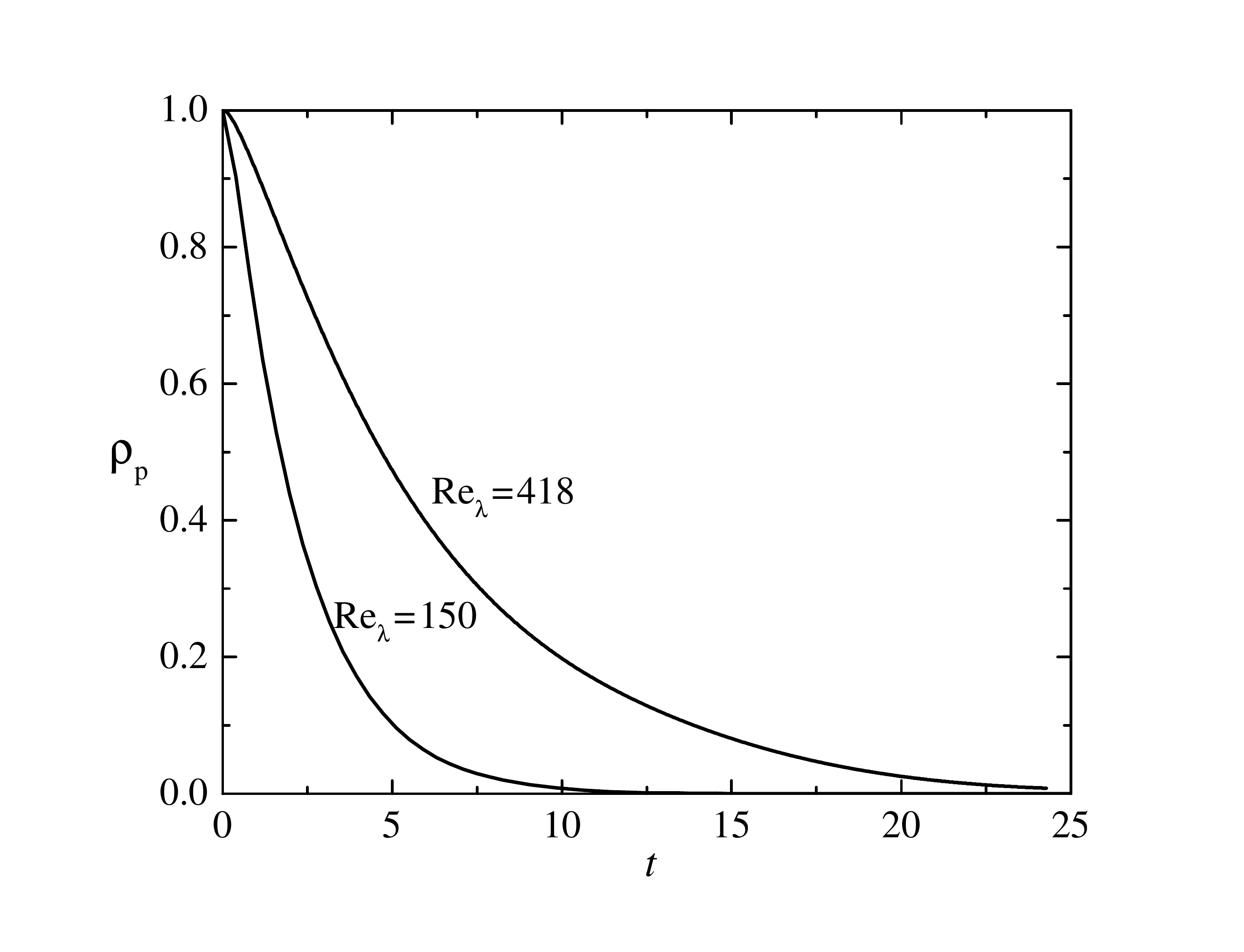}
		\caption{Lagrangian pressure autocorrelation functions at $\Re_\lambda=150$ and $418$.
                }\label{fig:presautocorr}
	\end{figure}

The results for $\mbox{Var}(p)$ are similar to those of Gotoh \& Rogallo \cite{gr99_jfm}\footnote{The relationship between the non-dimensional variables chosen by Gotoh \& Rogallo and those adopted here are: $F_p=(4/9)$ Var$(p)$.}, who report a value of about 0.9 at $\Re_\lambda=39$ which decreases to about 0.7 at $\Re_\lambda=172$. Cao et al \cite{ccd99_pf} report $\mbox{Var}(p)=0.883$ at $\Re_\lambda=131$, while Pumir \cite{pumir94_pf} obtained $\mbox{Var}(p)\simeq 1$ for $21.6\leq \Re_\lambda \leq 77.5$. Lagrangian Taylor and integral timescales for pressure are not available in the literature, nor is the Lagrangian pressure autocorrelation function, which is shown in Figure \ref{fig:presautocorr}.

The pressure PDFs exhibit the (approximately) exponential tails previously reported by Pumir \cite{pumir94_pf} and Cao et al \cite{ccd99_pf}. Specifically, the data can be fitted by
\begin{equation}
  \mbox{PDF}(p) \simeq \begin{cases}
    0.18\,\exp (1.1\,p) & \mbox{for }\,\Re_\lambda=150,\,-13\leq p \leq -5 \\
    0.008\,\exp (0.65\,p) & \mbox{for }\,\Re_\lambda=418,\,-17\leq p \leq -7
  \end{cases}
  \label{eqpdf}
\end{equation}


\subsection{Average frequency of pressure fluctuation events}

\begin{figure}[htb]
\def\svgwidth{\textwidth}
\centering
		\includegraphics[scale=0.35]{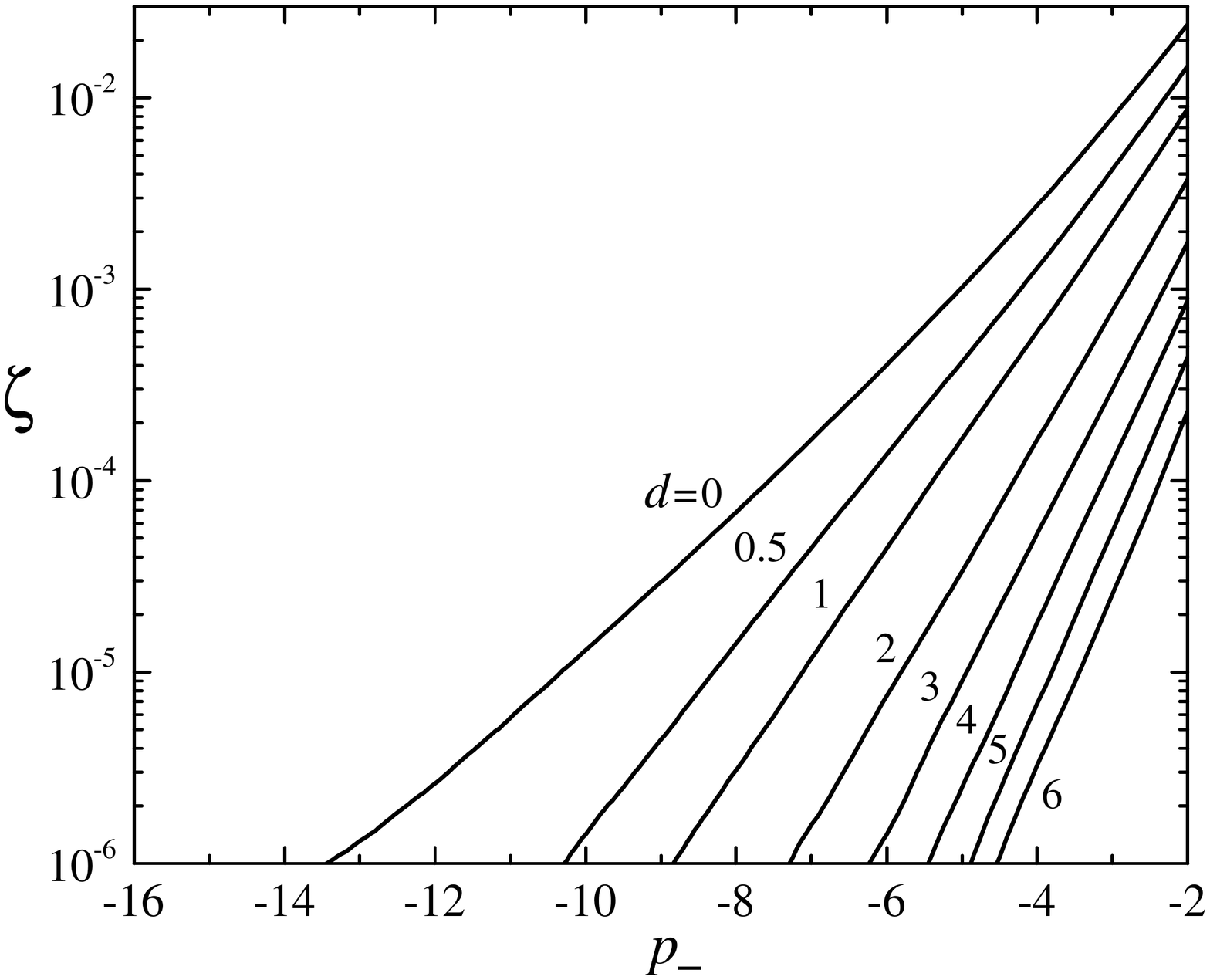}\\$~$(a)$~$\\
		\includegraphics[scale=0.35]{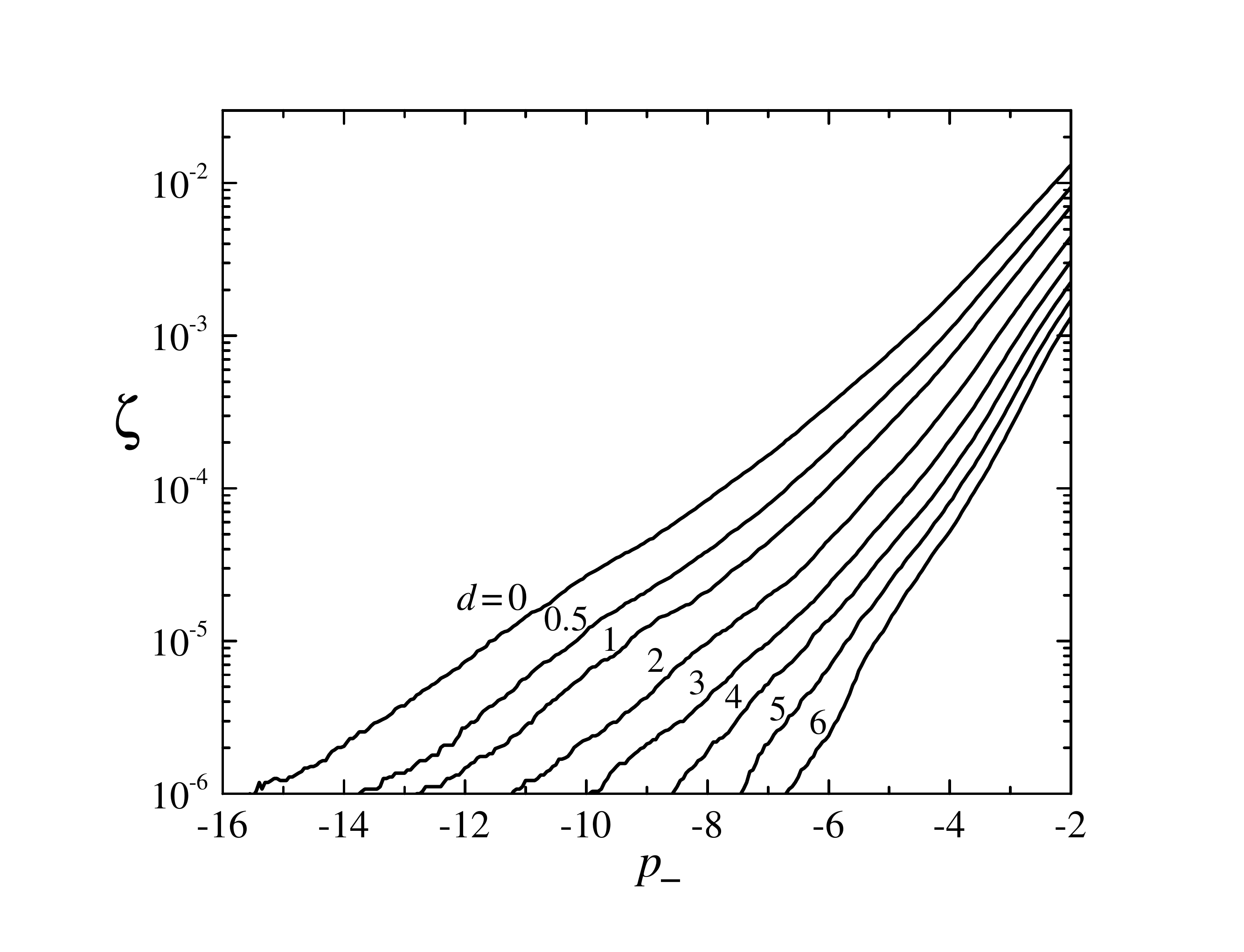}\\$~$(b)$~$
\caption{Non-dimensional frequency $\zeta(p_-,d)$ of low-pressure events as function of the threshold pressure $p_-$ for several values of the minimum duration $d=0, 0.5, 1, \ldots$. Results at (a) $\Re_\lambda=150$ and (b) $\Re_\lambda=418$. }\label{fig:zetaptsimple}
\end{figure}

The main novel results in this contribution are the values of $\zeta(p_-,d)$ as depicted in Figure \ref{fig:zetaptsimple}. Notice that $\zeta$ is an average frequency of events {\em per particle}. { An estimate of its relative statistical error can be obtained as $1/\sqrt{n(p_-,d,T)}$ for each computed value. This makes the relative error of $\zeta$ to be proportional to $\zeta^{-1/2}$, with a proportionality constant of $3\times 10^{-5}$ for the $\Re_\lambda=150$ simulation, and of $2\times 10^{-4}$ for the $\Re_\lambda=418$ one. No values of $\zeta$ below $10^{-6}$ are reported because for this value the relative statistical error is already at 20\% for the higher $\Re$. }


A typical use of these data would be as follows: Consider a volume $V$ of fluid in a (homogeneous, isotropic) turbulent flow characterized by given values of $u'$, $\lambda$ and $\Re_\lambda$, and assume that the fluid contains some concentration $Z$ of small particles (per m$^3$) in suspension. Then $F=ZV\zeta(p_-,d)u'/\lambda$ is the expected frequency (in events/sec) with which the suspended particles will undergo negative pressure fluctuations below $\rho u'^2 p_-$ (in Pa) that last more than $d\lambda/u'$ (in sec) within the volume $V$. Of course for this to hold the quantity $\zeta(p_-,d)$ must be evaluated at the flow's $\Re_\lambda$.

\begin{figure}[htb]
\def\svgwidth{\textwidth}
\centering 
		\includegraphics[scale=0.35]{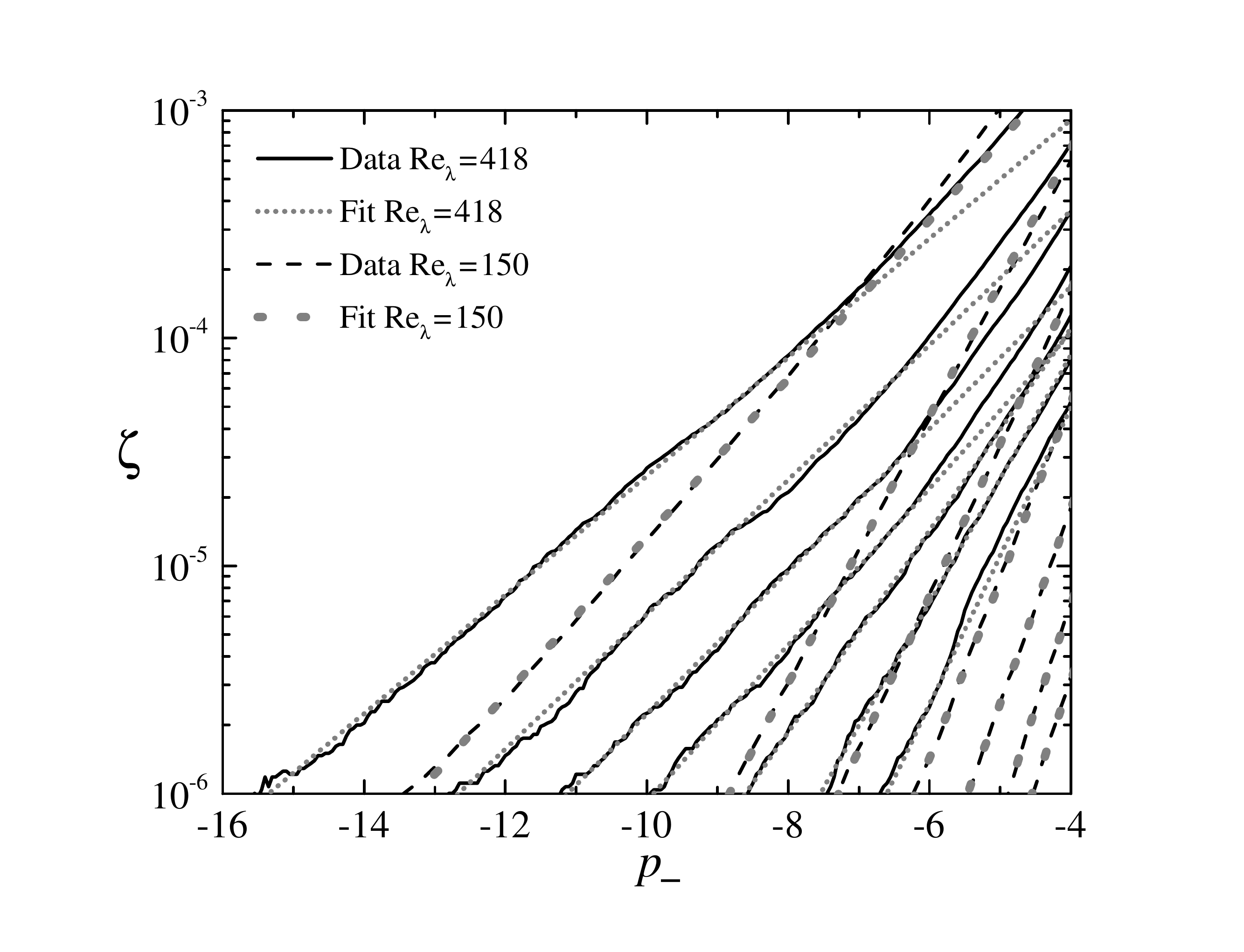}
\caption{$\zeta(p_-,d)$ for $d=0,1,\ldots,6$, for both values of $\Re_\lambda$, and their exponential fits. }
\label{fig:frequenciessimple}
\end{figure}

Going back to Figure \ref{fig:zetaptsimple}, notice that for any minimum duration $d$ the average frequency is approximately exponential in $p_-$, i.e., $\zeta(p_-)\simeq C\,\exp (\beta\,p_-)$, with $C$ and $\beta$ depending on $d$ and on $\Re_\lambda$. In fact, some upward concavity can be observed in the semilog plots for the higher values of $p_-$ (milder pressure fluctuations), but for lower values the plots closely follow straight lines. This can be more easily seen in Fig. \ref{fig:frequenciessimple}, in which all plots of $\zeta(p_-,d)$ corresponding to $d=0,1,2,\ldots,6$ for both values of $\Re_\lambda$ have been put together simultaneously with their exponential fits. 


\begin{table}
\begin{center}
\begin{tabular}{lcccc}
  \hline
   & \multicolumn{2}{c}{$\Re_\lambda=150$} & \multicolumn{2}{c}{$\Re_\lambda=418$} \\ \hline
  Min. duration ($d$) & $~~~~C~~~~$ & $~~~~\beta~~~~$ & $~~~~C~~~~$ & $~~~~\beta~~~~$ \\ \hline
  0 & 0.04 & 0.80 & 0.010 & 0.60 \\
  1 & 0.15 & 1.35 & 0.0055 & 0.68 \\
  2 & 0.08 & 1.55 & 0.003 & 0.72 \\
  3 & 0.07 & 1.80 & 0.0025 & 0.79 \\
  4 & 0.047 & 1.97 & 0.0065 & 1.02 \\
  5 & 0.03 & 2.10 & 0.0125 & 1.25 \\
  6 & 0.014 & 2.10 & 0.022 & 1.50 \\ \hline
\end{tabular}
\end{center}
\caption{Constants $C$ and $\beta$ of the exponential fits of the data, for low-pressure events of duration greater or equal than $d$ and for the two datasets ($\Re_\lambda=150$ and 418). The quality of the fits can be appreciated in Fig. \ref{fig:frequenciessimple}.}\label{tablecbeta}
\end{table}

The fitted values of $C$ and $\beta$ are listed in Table \ref{tablecbeta}. Most relevant is the parameter $\beta$, i.e., the slope in the semi-log plots. A smaller $\beta$ implies that, as the threshold is lowered, the average rate of excursions below the threshold decreases more slowly. For any minimum duration of the excursions, $\beta$ is smaller for the higher Reynolds number. This is consistent with the pressure PDFs, which also show that lower pressures become more probable at higher $\Re_\lambda$, with exponential tails that behave as $\exp (\gamma p_-)$, with $\gamma=1.1$ at $\Re_\lambda=150$ and $\gamma=0.65$ at $\Re_\lambda=418$ (see Fig. \ref{fig:pressurepdfs}).

By direct inspection of Table \ref{tablecbeta} one notices that $\gamma$ is, for both $\Re_\lambda$, close to the parameter $\beta$ that corresponds to events of minimum duration $d$ between $0$ and $1$ (i.e., between zero and roughly four Kolmogorov time scales). The experimental consequences of this are quite interesting. Assume that, as proposed by LaPorta et al \cite{lvmb00_pf}, one uses the gas nuclei in a liquid as pressure sensors and measures the average rate of cavitation events as a function of the (variable) reference pressure $p_{\mbox{\scriptsize{ref}}}$ of the flow. Then, {\em if the nuclei's reaction time is smaller than a few Kolmogorov time scales}, the curve {\em cavitation rate vs. $(\pcav - p_{\textnormal{\scriptsize{ref}}})$} will be approximately proportional to $\int_{-\infty}^{\pcav-\pref} \mbox{PDF}(p)\,dp$ and thus, since PDF$(p)$ is exponential, to PDF$(\pcav-\pref)$ itself. This has been argued in the literature by several authors \cite{ag79,rk94_jfm,lvmb00_pf}  on the basis of physical intuition and is herein rigorously confirmed. Notice that the short reaction time of the nuclei is crucial for the proportionality $\zeta(p_-)\sim \mbox{PDF}(p_-)$ to hold. If the reaction time is larger than, say, 20 $\tau_K$ (corresponding to $d\simeq 5$), the dependence of the cavitation rate with pressure will largely differ from that of the pressure PDF. It is worth mentioning that the values of $\beta$ inferred by LaPorta et al \cite{lvmb00_pf}, which were in the range 0.14-0.22 for $\Re_\lambda$ between 1658 and 1880, suggest that $\beta$ keeps decreasing with $\Re_\lambda$ in much the same way as observed from the DNS databases studied here.

\begin{figure}[htb]
\centering 
\def\svgwidth{\textwidth}
		\includegraphics[scale=0.35]{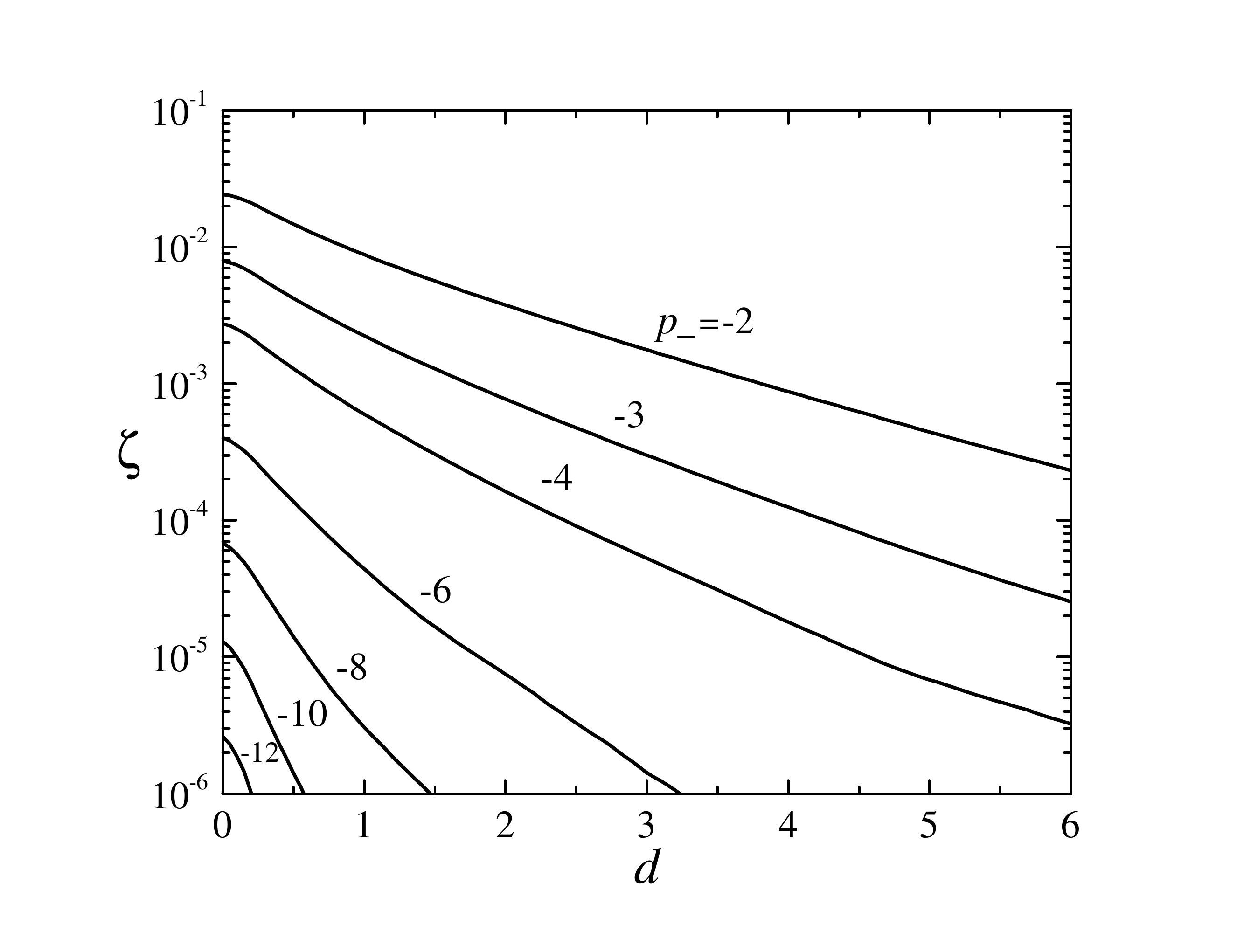}\\$~$(a)$~$\\
		\includegraphics[scale=0.35]{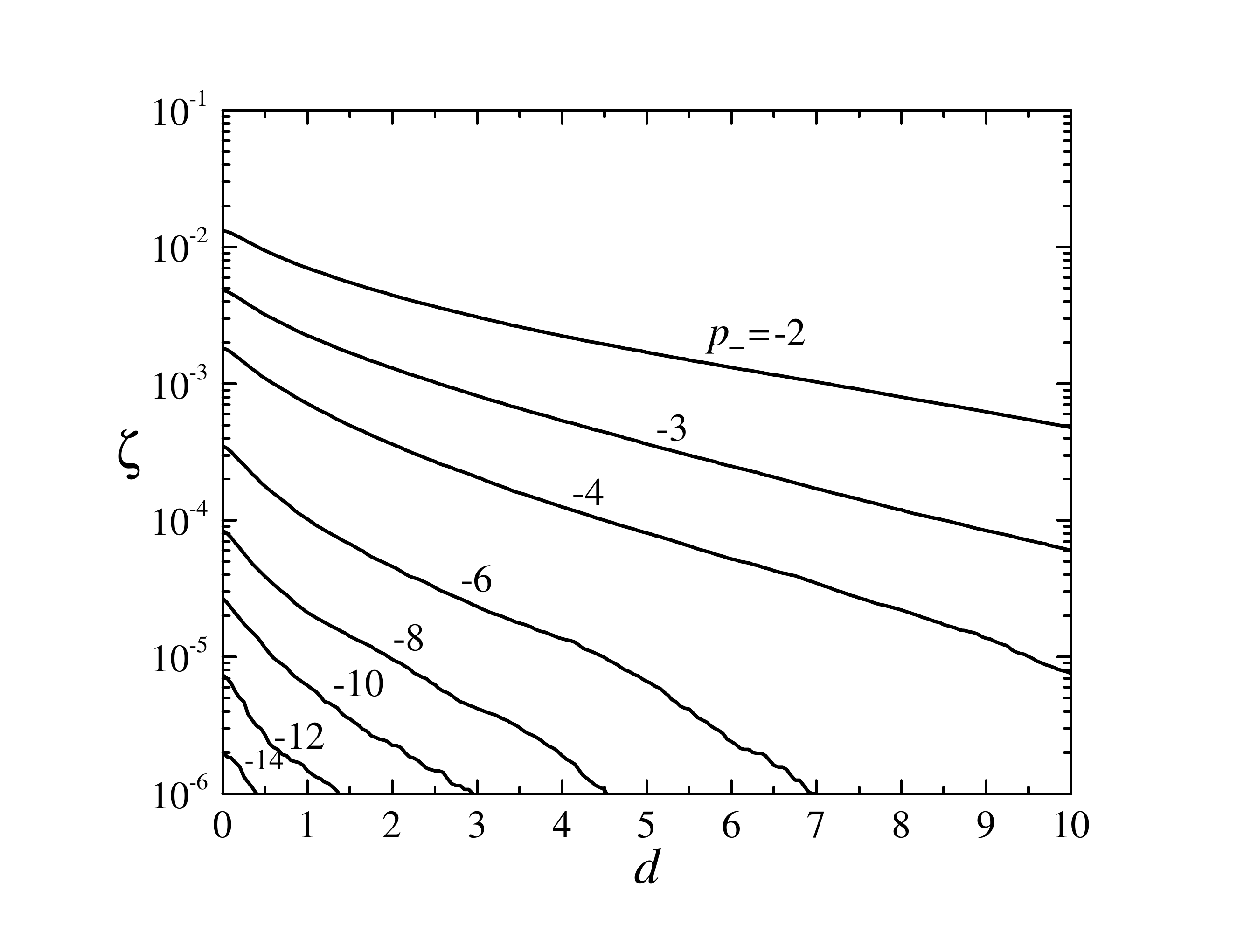}\\$~$(b)$~$
\caption{Non-dimensional frequency $\zeta(p_-,d)$ of low-pressure events as function of minimum duration $d$ for several values of the threshold pressure ($p_-=-2, -3, -4, \ldots$). Results at (a) $\Re_\lambda=150$ and (b) $\Re_\lambda=418$.}\label{fig:zetatpsimple}
\end{figure}

The dependence of $\zeta$ with the minimum duration $d$ of the events is depicted in Fig. \ref{fig:zetatpsimple}. An approximately exponential decay of $\zeta$ with $d$ is observed for $d$ greater than a few $\tau_K$ (say, $d>0.5$), with a logarithmic slope that becomes more negative as the threshold $p_-$ is lowered. For any $p_-$ the average rate of long events increases significantly with $\Re_\lambda$.

\begin{figure}[htb]
\centering 
\def\svgwidth{\textwidth}	
		\includegraphics[scale=0.35]{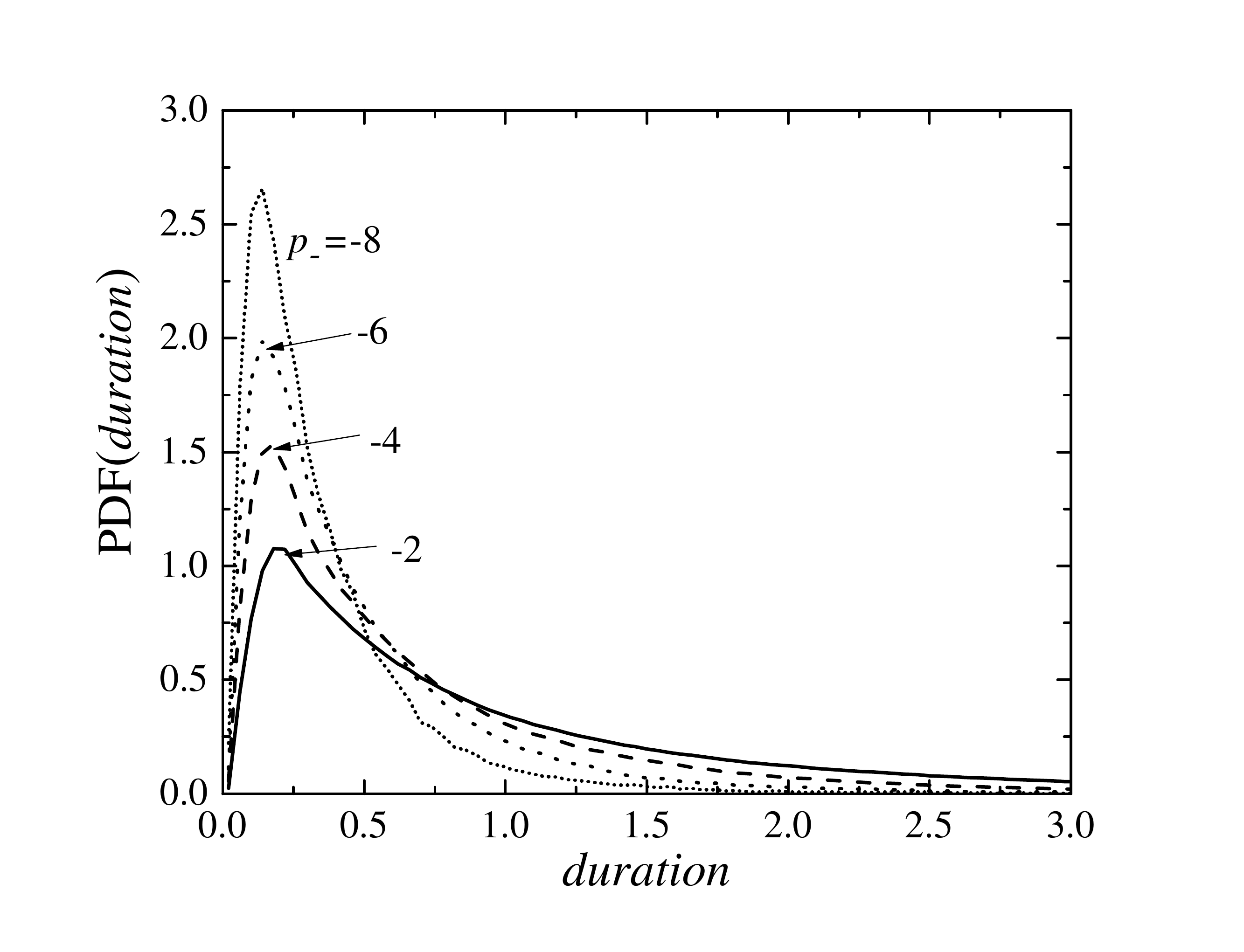}\\$~$(a)$~$\\
		\includegraphics[scale=0.35]{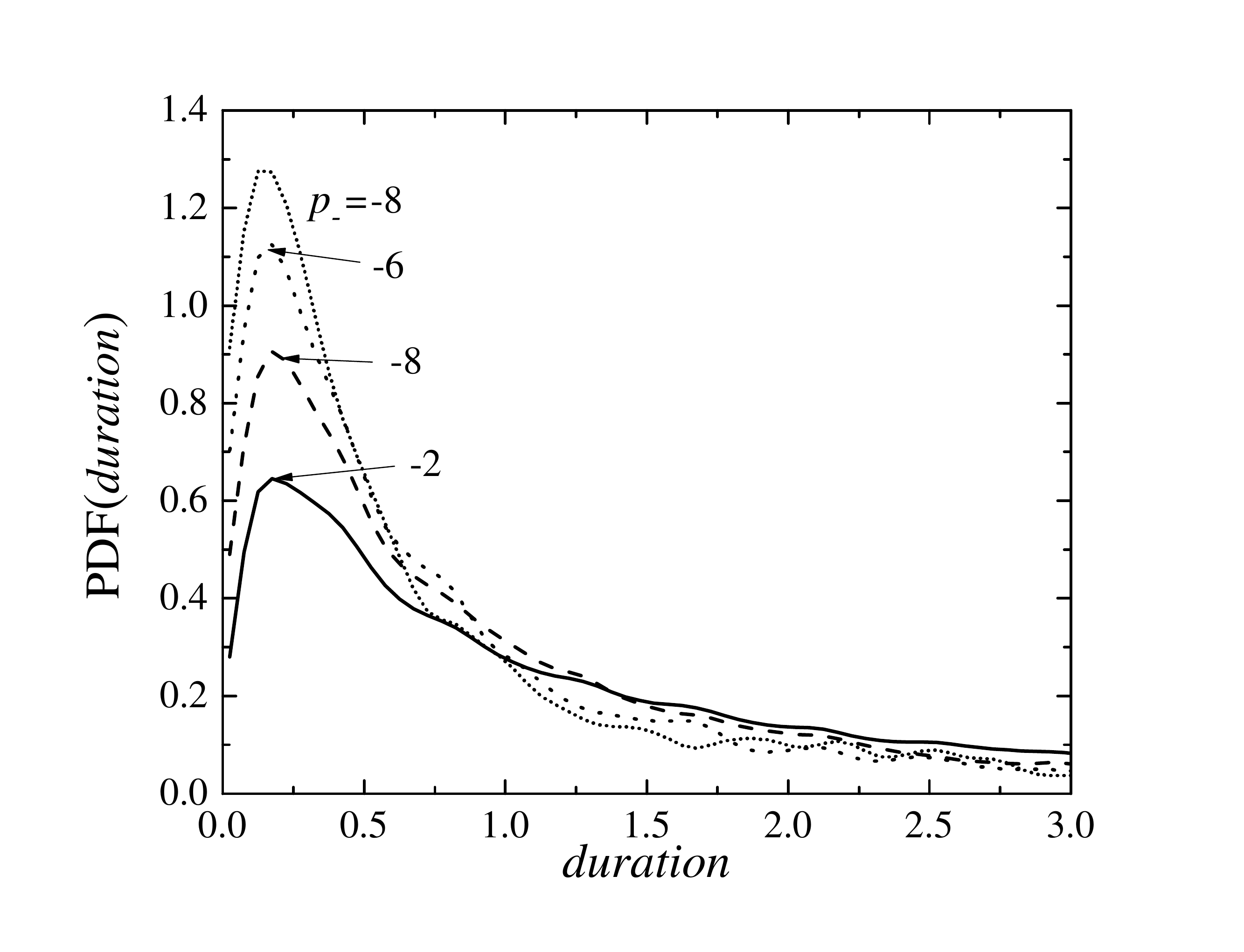}\\$~$(b)$~$
\caption{PDFs of the duration of the low-pressure events for several values of the threshold pressure ($p_-=-2, -4, -6, -8$). Results at (a) $\Re_\lambda=150$ and (b) $\Re_\lambda=418$.}\label{fig:pdfduration}
\end{figure}

\begin{figure}[htb]
\centering 
\def\svgwidth{\textwidth}	
		\includegraphics[scale=0.35]{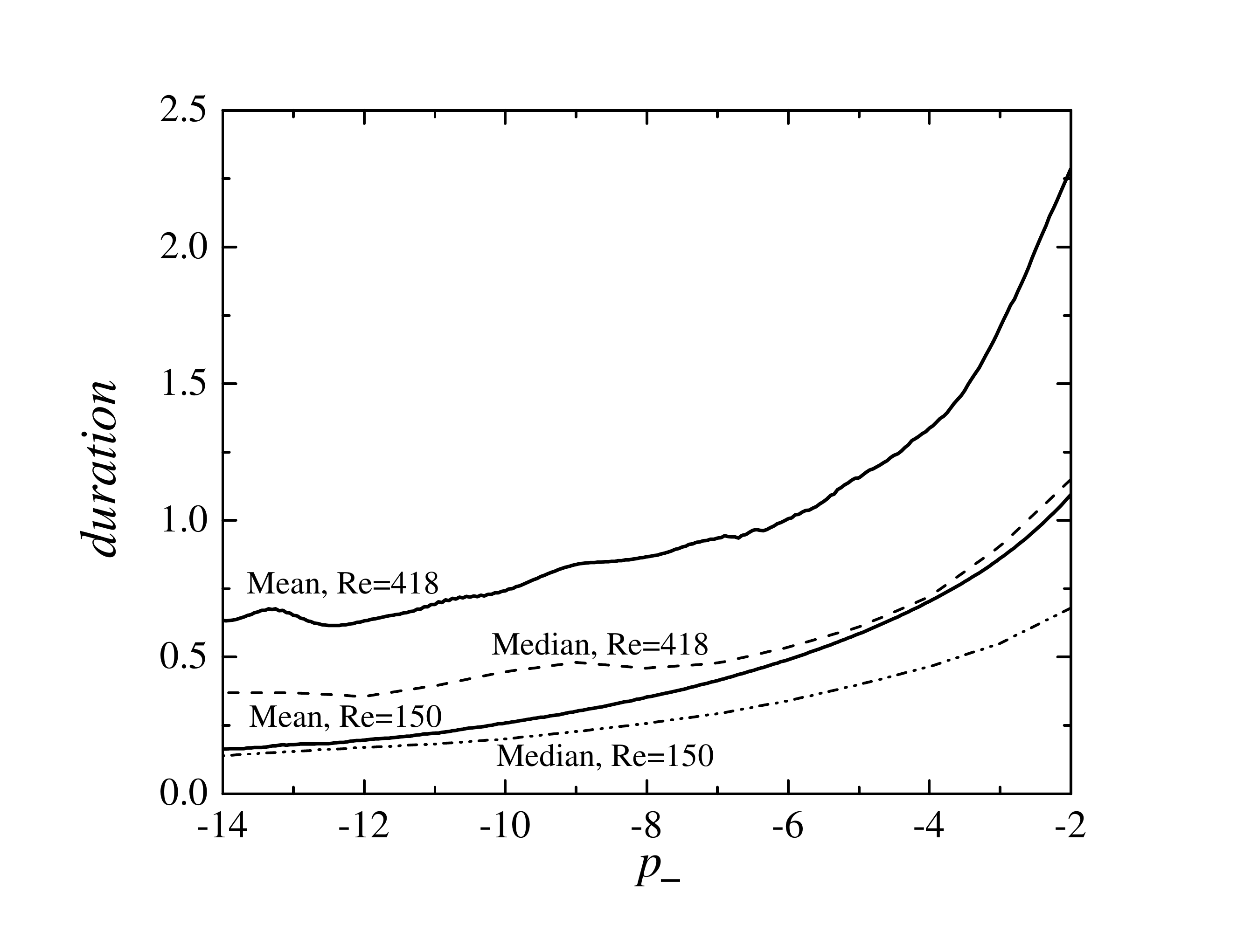}
\caption{Mean and median duration of the low-pressure events as functions of the threshold pressure $p_-$.}\label{fig:meansandmedians}
\end{figure}

It is informative to complement the previous results with further analysis of the distribution of the duration ($t_{\mbox{\scriptsize{end}}}-t_{\mbox{\scriptsize{start}}}$) of the events corresponding to each threshold $p_-$. The corresponding PDFs are shown in Fig. \ref{fig:pdfduration}. It is observed that the peak of the PDF is always about 2/3 of $\tau_K$ and that the PDFs for each $p_-$ have longer tails for the larger $\Re_\lambda$.  As $p_-$ decreases the distribution becomes narrower, i.e., events that last longer than $\tau_K$ become increasingly improbable. The means and medians of these PDFs are shown as functions of $p_-$ in Fig. \ref{fig:meansandmedians}. They increase significantly with $\Re_\lambda$, contrary to what happens with the peak of the duration PDFs.

\begin{figure}[htb]
\centering 
\def\svgwidth{\textwidth}
		\includegraphics[scale=0.35]{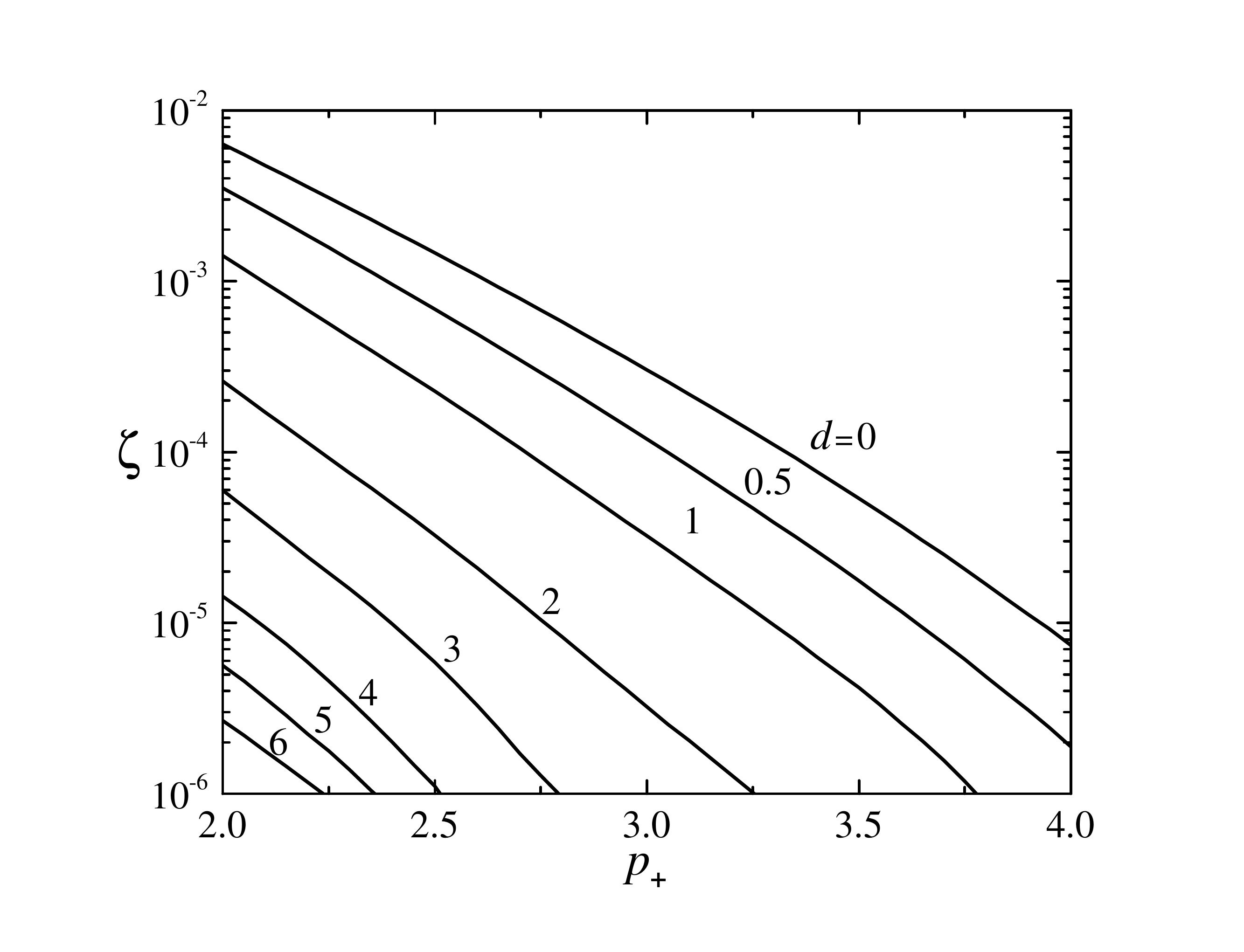}\\$~$(a)$~$\\
		\includegraphics[scale=0.35]{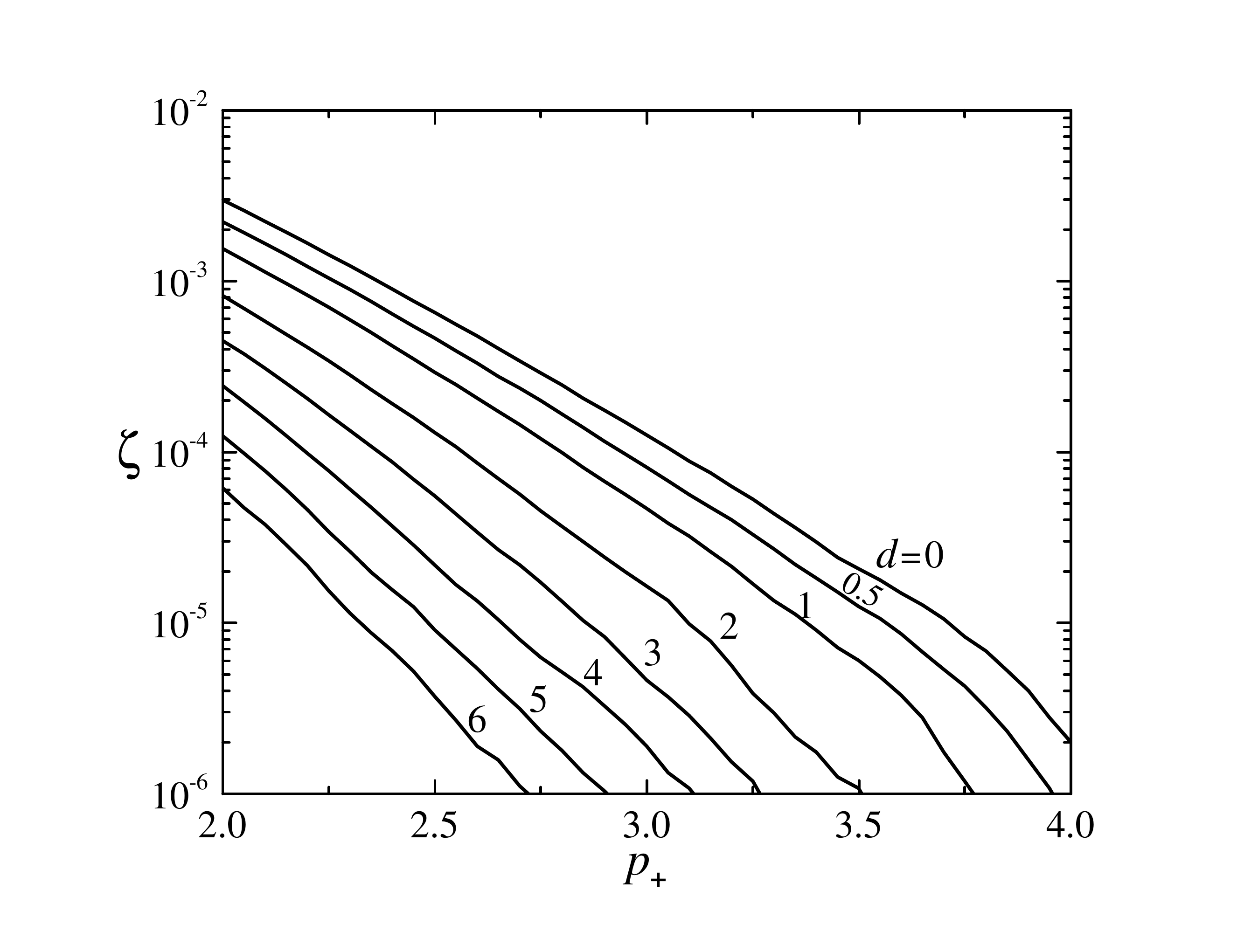}\\$~$(b)$~$
\caption{Obtained values of non-dimensional frequency $\zeta(p_+,d)$ of high-pressure events as function of the threshold pressure $p_+$ for several values of the minimum duration $d$. Results at (a) $\Re_\lambda=150$ and (b) $\Re_\lambda=418$. }\label{fig:zetapthighsimple}
\end{figure}

For the purpose of both completeness and comparison, let us close this section with plots of $\zeta(p_+,d)$, corresponding to {\em positive} pressure fluctuations. They are shown in Fig. \ref{fig:zetapthighsimple}. It is evident that positive pressure excursions are much more rare than negative ones, a result that is consistent with the rapid decay of the positive side of the pressure PDF. For minimum duration $d=0$ the rate of events does not change much with $\Re_\lambda$, but for events significantly longer than 10 Kolmogorov timescales ($d>2.58$) the non-dimensional frequency for $\Re_\lambda=418$ is much greater than that for $\Re_\lambda=150$, by an order of magnitude or more.

\subsection{Randomness of low-pressure fluctuations}

The occurrence of low-pressure fluctuations (of threshold $p_-$ and minimum duration $d$) in a turbulent flow is certainly a random process. Its stochastic properties can be investigated looking at the {\em arrival process} formed by the (monotone increasing) sequence $t_{\mbox{\scriptsize{start}}}^{(1)},t_{\mbox{\scriptsize{start}}}^{(2)},\ldots$, where $t_{\mbox{\scriptsize{start}}}^{(i)}$ is the starting time of the $i$-th event. This generates the stochastic process of {\em interarrival times}, $D_i=t_{\mbox{\scriptsize{start}}}^{(i+1)}-t_{\mbox{\scriptsize{start}}}^{(i)}$. If $m$ random particles are seeded into the flow, then by definition the average interarrival time satisfies { $\overline{D}=1/(m\zeta(p_-,d))$}. A totally random arrival process (Poisson process) exhibits an exponential distribution for $D$, i.e.,
        \begin{equation}
          \mbox{PDF}(D)=\frac{1}{\overline{D}}\,\exp \left ( -\frac{D}{\overline{D}} \right ).
        \end{equation}
        Whether the sequence of low-pressure events is a Poisson process or not can thus be assessed by inspecting the PDF of interarrival times. This PDF cannot be built from the raw data consisting of all events undergone by the $M$ particles because of the temporal resolution limit imposed by the simulation time step. For example, for the case of events with $p_-=-2$ and $d=0$ the total number of such events recorded was { $2.11\times 10^7$} for the simulation at $\Re_\lambda=150$ ($M=10^6$) and { $3.43\times 10^5$} for that at $\Re_\lambda=418$ ($M=4\times 10^5$). Meanwhile, the number of simulated time steps is 20,000 and 5,000, respectively. This makes the interarrival times to be much smaller than the time step and thus poorly resolved.

        The procedure adopted to build $PDF(D)$ was as follows: From the $M$ particles in the simulation, batches of $m$ particles were extracted at random, selecting the number $m$ such that $\overline{D}$ equals 100 simulation time steps. For each batch the interarrival times were computed, and $\mbox{PDF}(D)$ was obtained averaging the histograms of 50,000 such batches.

\begin{figure}[htb]
		\centering 
		\def\svgwidth{\textwidth}	
		\includegraphics[scale=0.35]{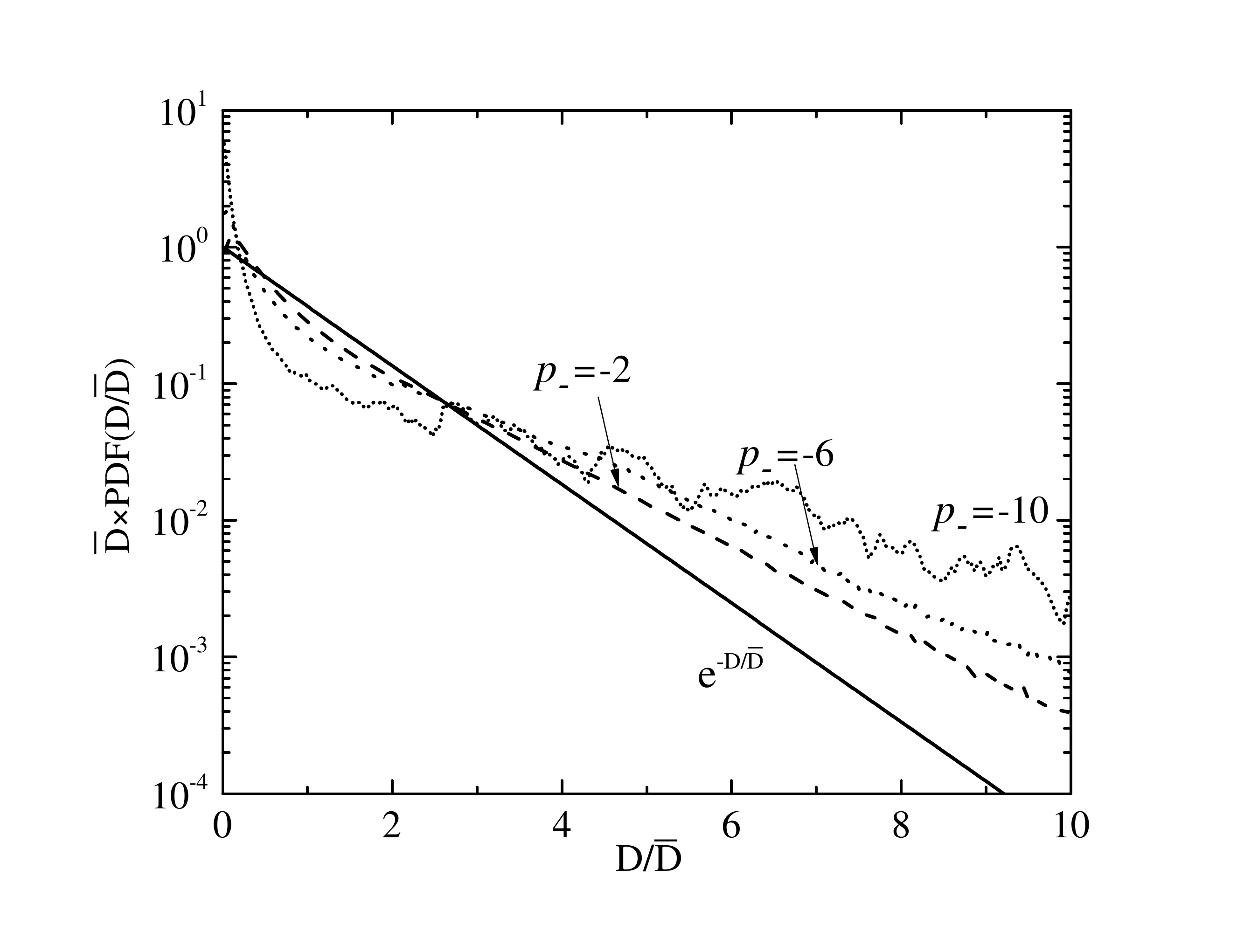}\\$~$(a)$~$\\
		\includegraphics[scale=0.35]{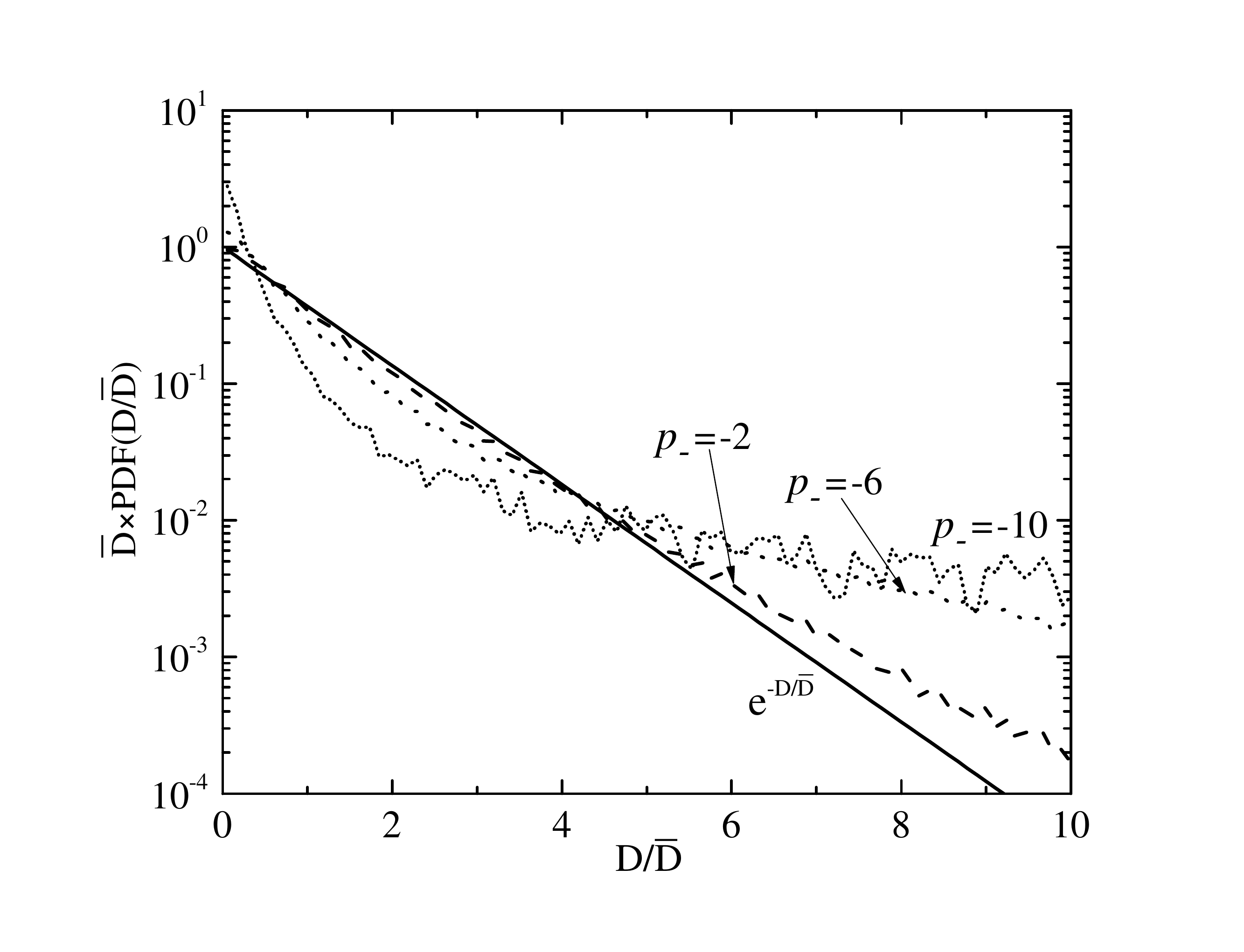}\\$~$(b)$~$
		\caption{PDF of the interarrival times of low pressure events of any duration ($d=0$) for different thresholds $p_-$. The exponential corresponding to a Poisson process is also plotted for comparison. Results at (a) $\Re_\lambda=150$ and (b) $\Re_\lambda=418$. }\label{fig:pdfdelaysminus}
	\end{figure}

The results considering events of any duration ($d=0$) for each pressure threshold are shown in Fig. \ref{fig:pdfdelaysminus}, where we plot $\overline{D}\,\mbox{PDF}(D)$ as a function of $D/\overline{D}$. Also shown is the exponential distribution $\exp(-D/\overline{D})$ which would correspond to a Poisson process. Analogous PDFs can be built for other values of $d$, but they are not shown since they are quite similar. 

\begin{figure}[htb]
\centering 
\def\svgwidth{\textwidth}	
		\includegraphics[scale=0.35]{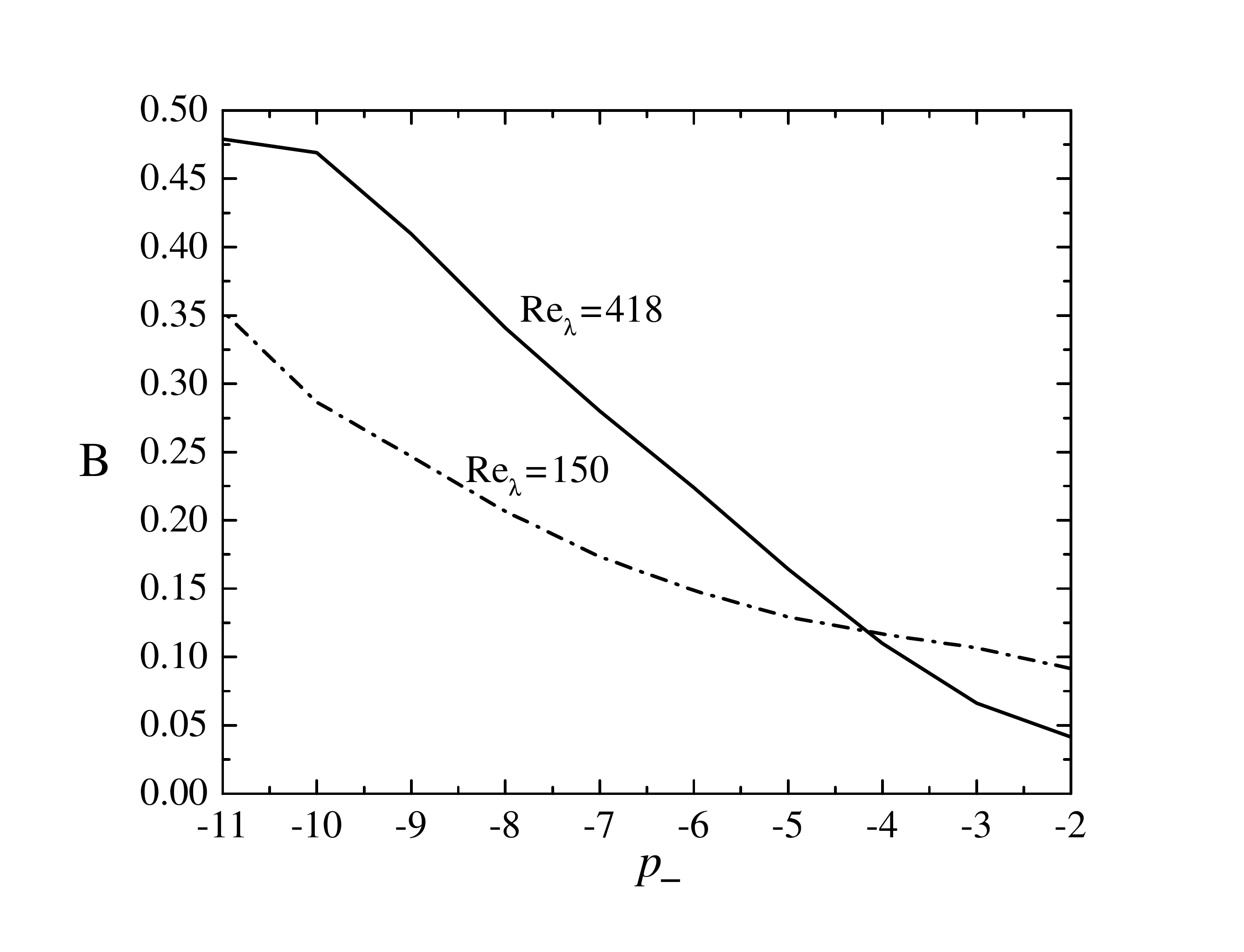}
\caption{Burstiness parameter of the low-pressure events as a function of the threshold pressure $p_-$ for the two $\Re_\lambda$ considered.}\label{fig:statsburstiness}
\end{figure}

\begin{figure}[htb]
\centering 
\def\svgwidth{\textwidth}	
		\includegraphics[scale=0.35]{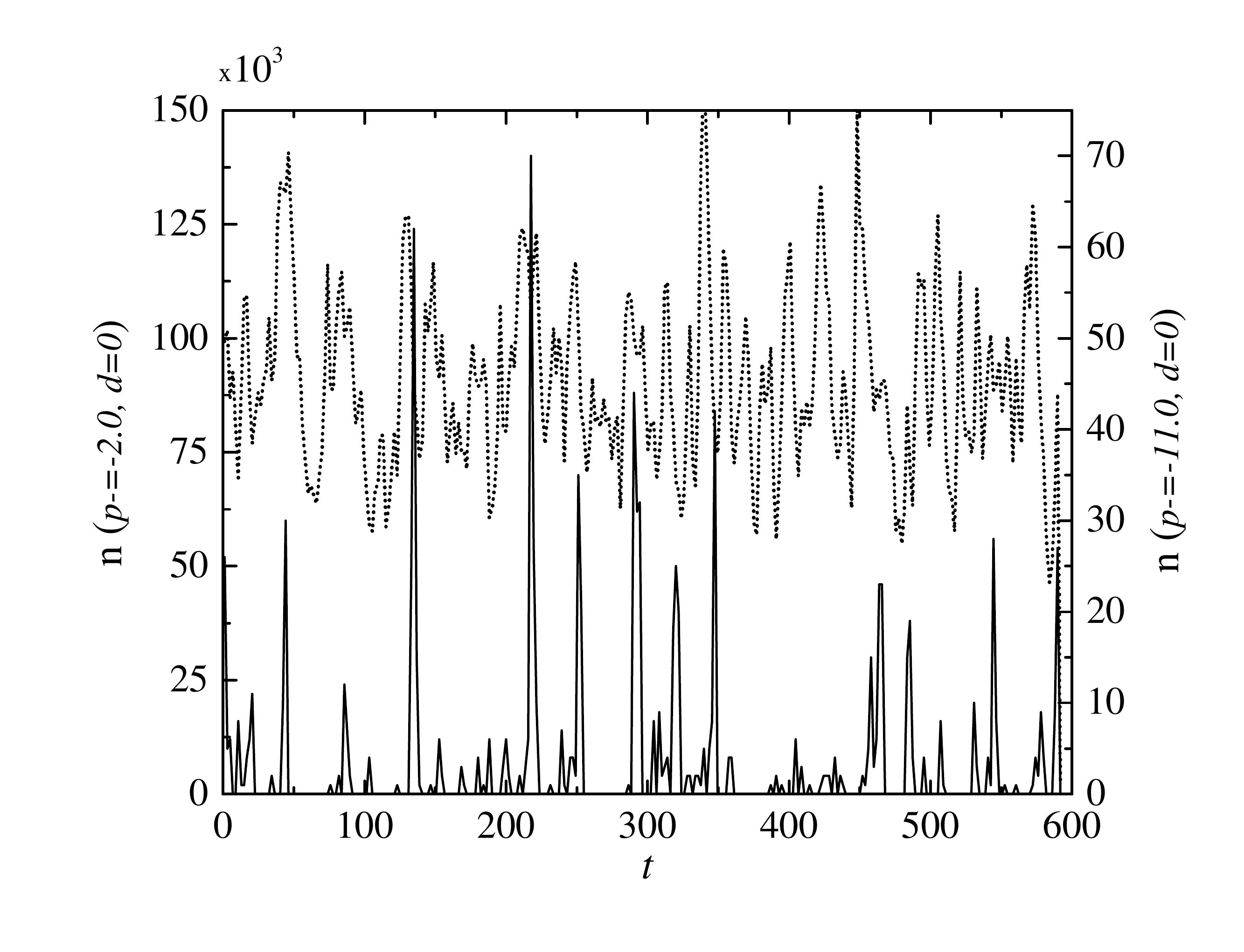}
\caption{Number of events that start within a time bin (size of the bin: 50 time steps) as a function of time. The plots correspond to events of threshold $p_-=-2$ (dotted line) and -11 (solid line), with minimum duration $d=0$, as recorded in the $\Re_\lambda=150$ simulation. }\label{fig:bursts}
\end{figure}

        The first immediate observation from the plots of PDF$(D)$ is that low-pressure events do not take place as a totally random, Poisson process that would entail an exponential PDF. The semilog-plots of the PDFs of interarrival times exhibit an upward concavity, or ``heavy tail'', which becomes more prominent as the threshold $p_-$ is decreased. This heavy tail is characteristic of processes that exhibit {\em burstiness}, for which a popular quantitative measure in the literature is the ``burstiness parameter'' \cite{gb08_epl} defined by
        \begin{equation}
          B = \frac{\sigma(D)-\overline{D}}{\sigma(D)+\overline{D}},
        \end{equation}
        where $\sigma(D)$ is the standard deviation of $D$. Notice that $B=-1$ for a periodic process ($\sigma(D)=0$),  $B=0$ for a Poisson process ($\sigma(D)=\overline{D}$), and $B=1$ for a highly bursty process ($\sigma(D) \gg \overline{D}$). The graphs of $B$ {\em vs.} $p_-$ fixing the minimum duration to $d=0$ are shown in Fig. \ref{fig:statsburstiness} for both $\Re_\lambda$ considered. Clearly, the low-pressure events are more bursty for larger fluctuations (more negative $p_-$). To visualize this, in Fig. \ref{fig:bursts} we plot, as a function of time, the number of events that start in temporal bins of 50 time steps for $p_-=-2$ and $p_-=-11$ in the simulation with $\Re_\lambda=150$. The number of events per bin with threshold -2 oscillates moderately around its mean value, while that with threshold -11 is most of the time near zero with intermittent bursts that reach 30 or more events per bin. It is remarkable that burstiness parameters of value 0.3 and higher are observed.  Such values are not frequent in natural phenomena, and can be found in highly intermittent human activities such as e-mail sending \cite{gb08_epl}.

\begin{figure*}[htb]
\centering 
\def\svgwidth{\textwidth}	
\includegraphics[scale=.075]{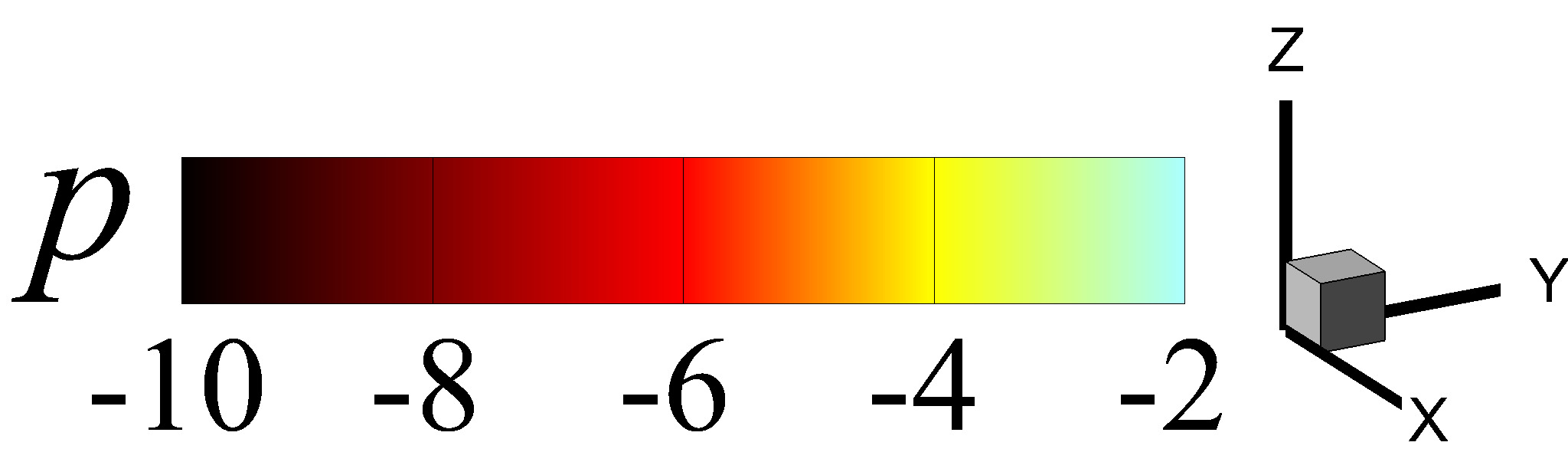}\\
\includegraphics[scale=.076]{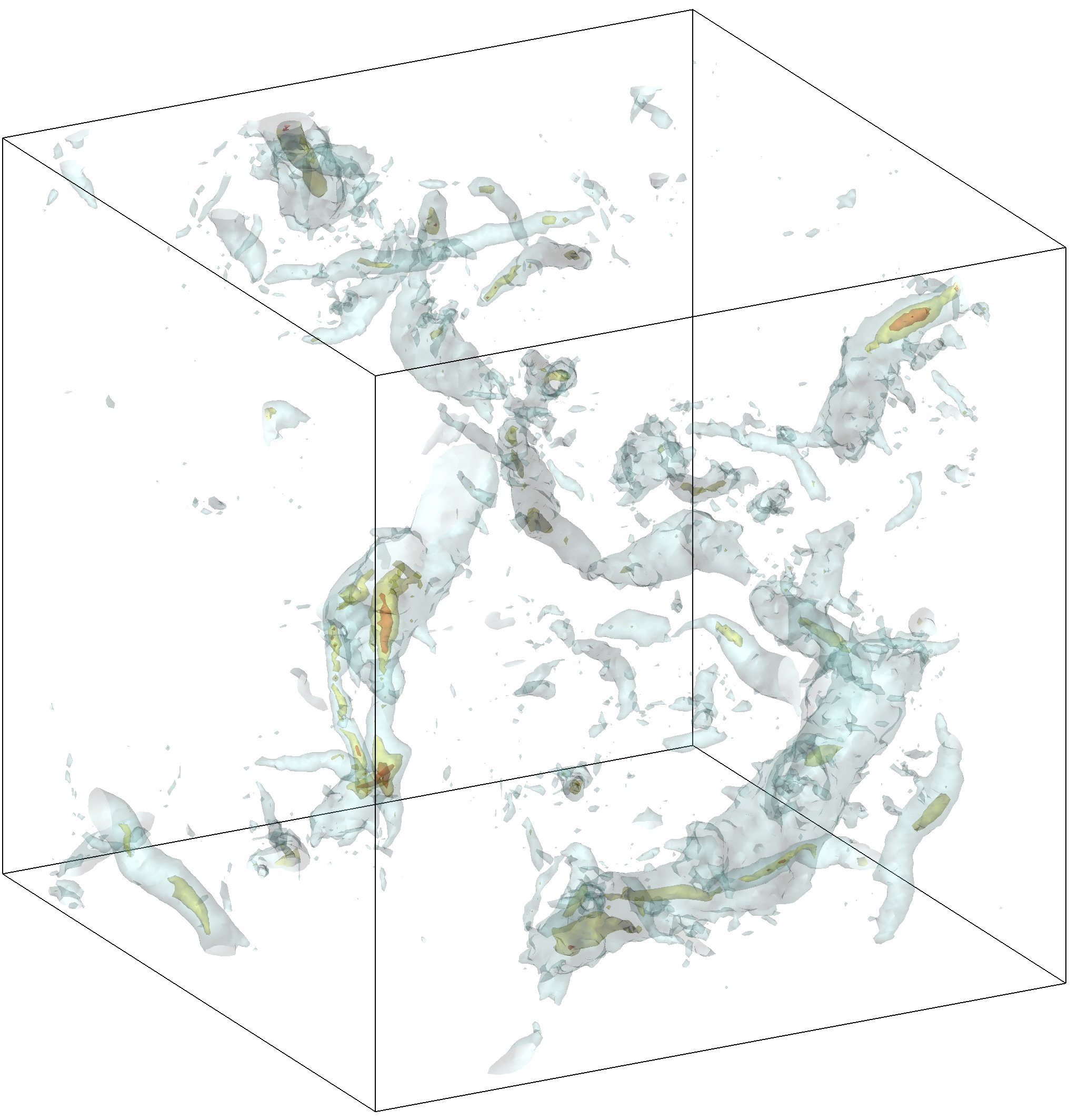}
\includegraphics[scale=.076]{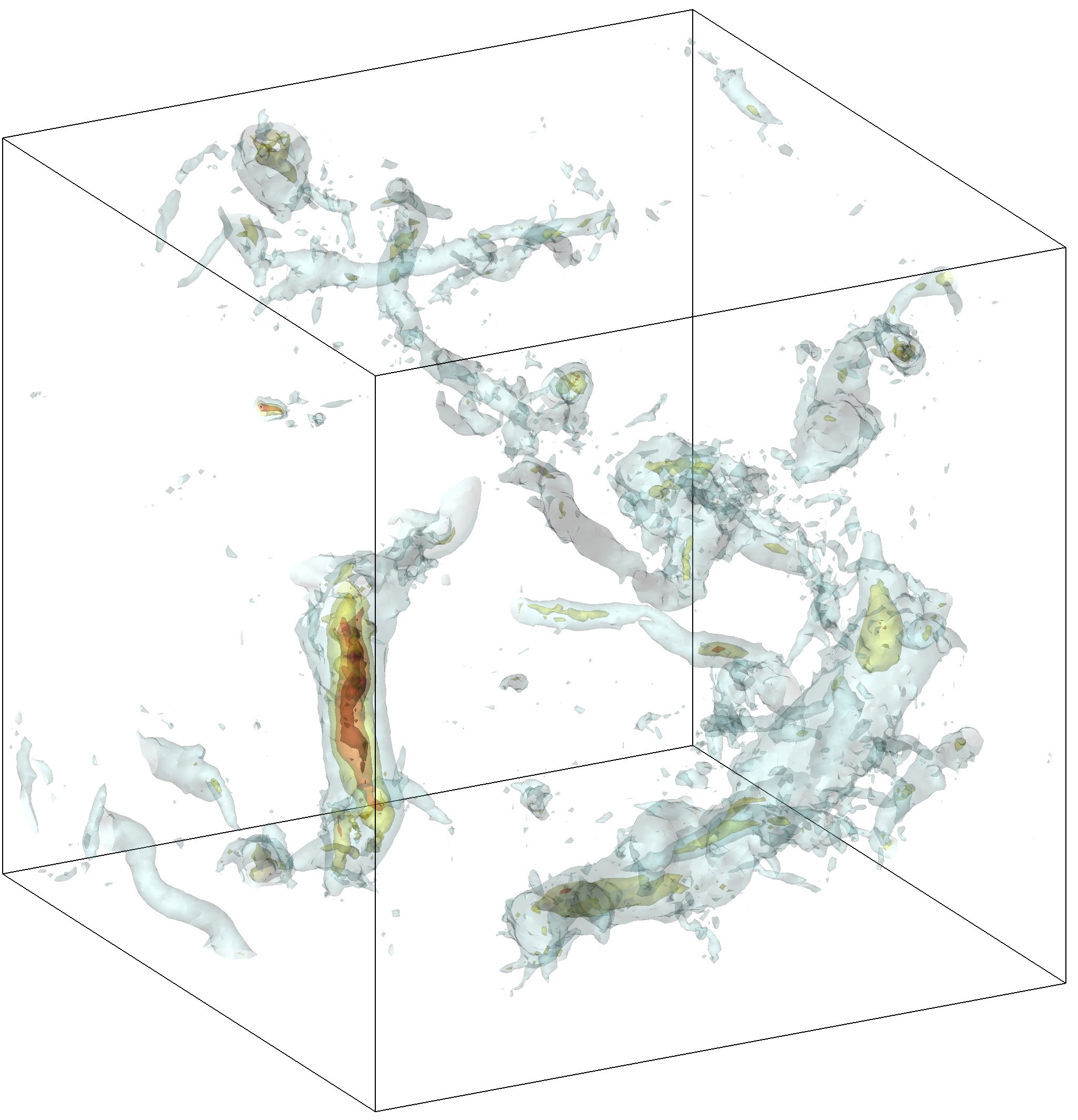}
\includegraphics[scale=.076]{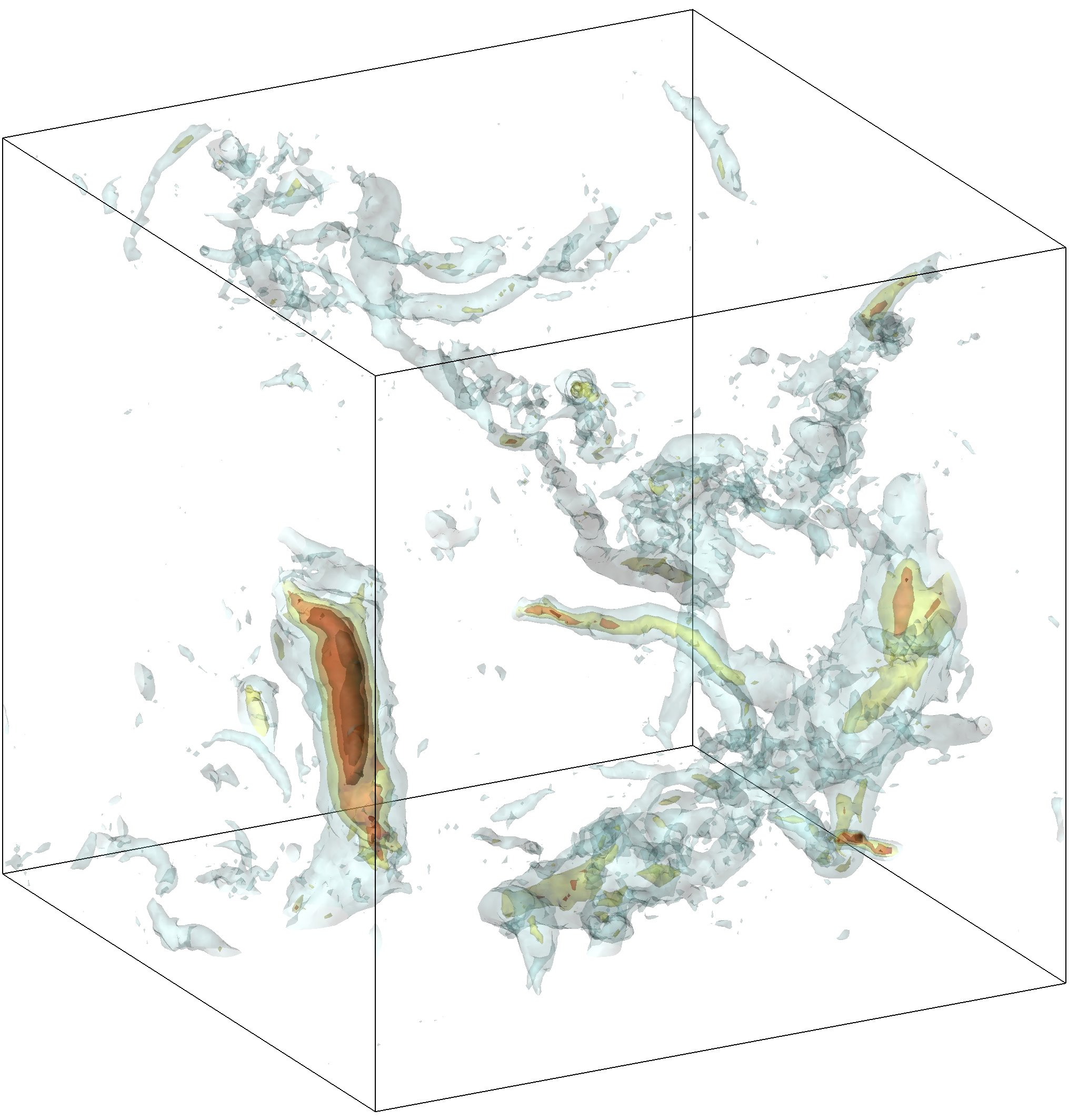}
\includegraphics[scale=.076]{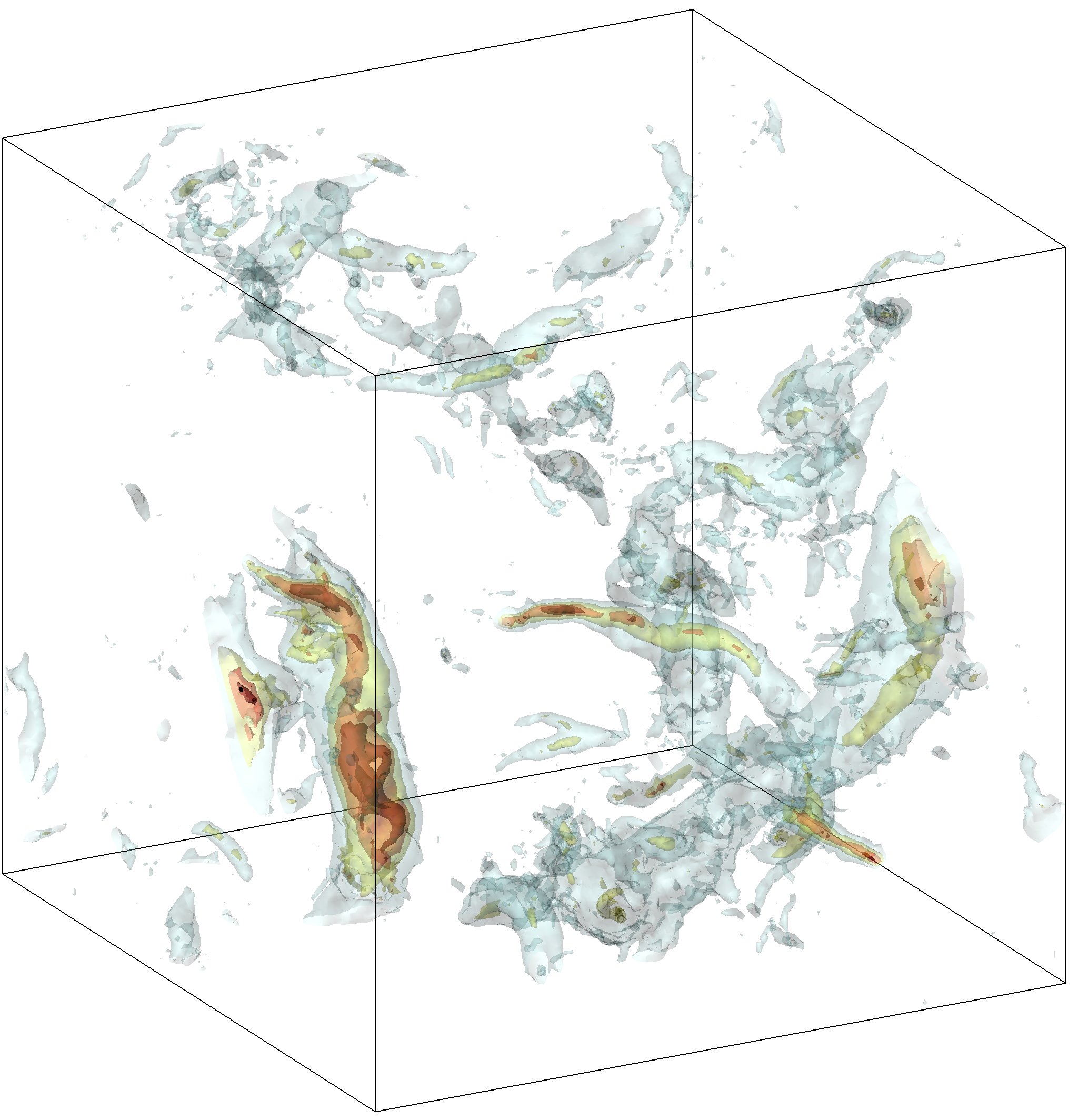}
\includegraphics[scale=.076]{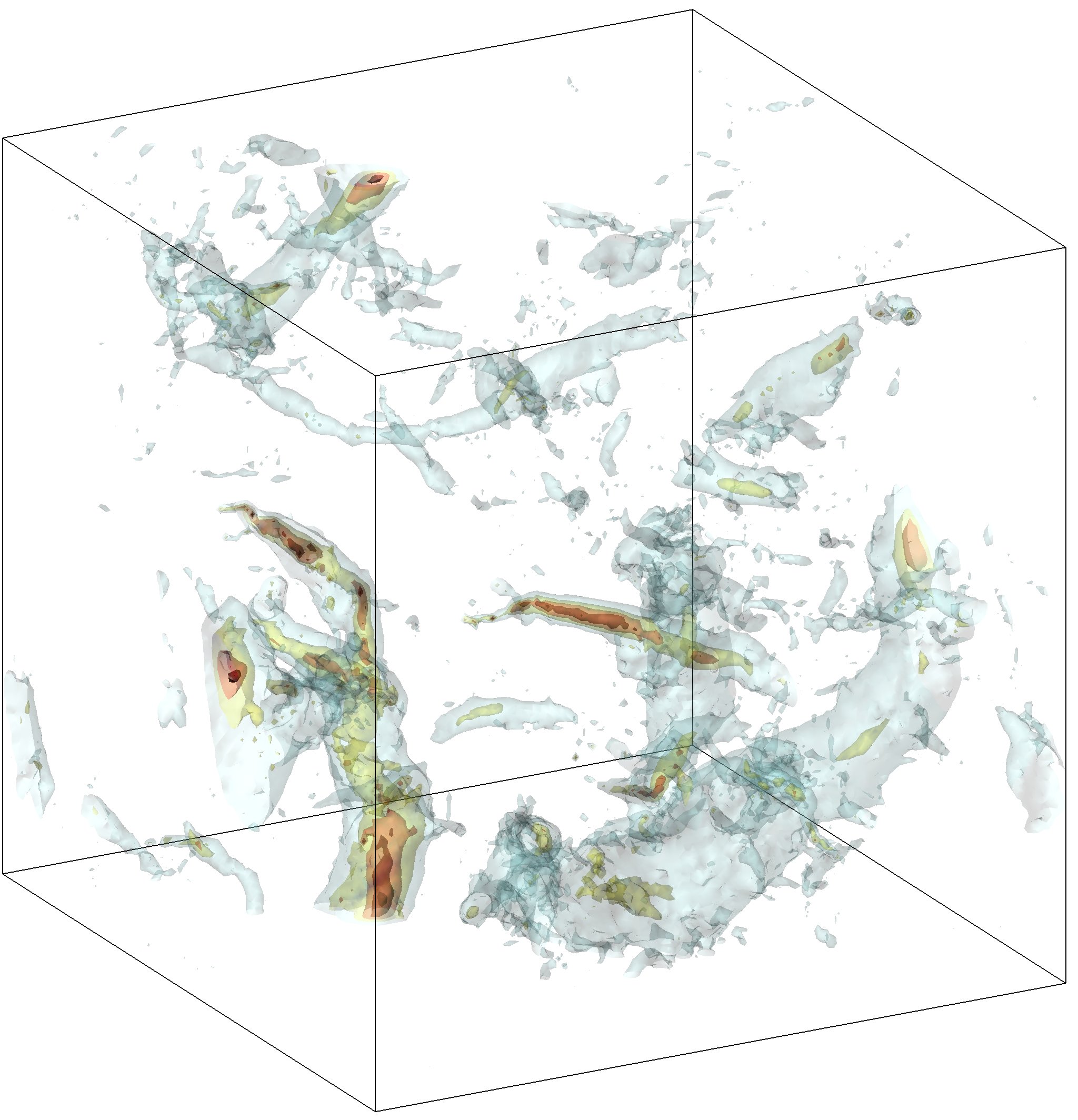}
\includegraphics[scale=.076]{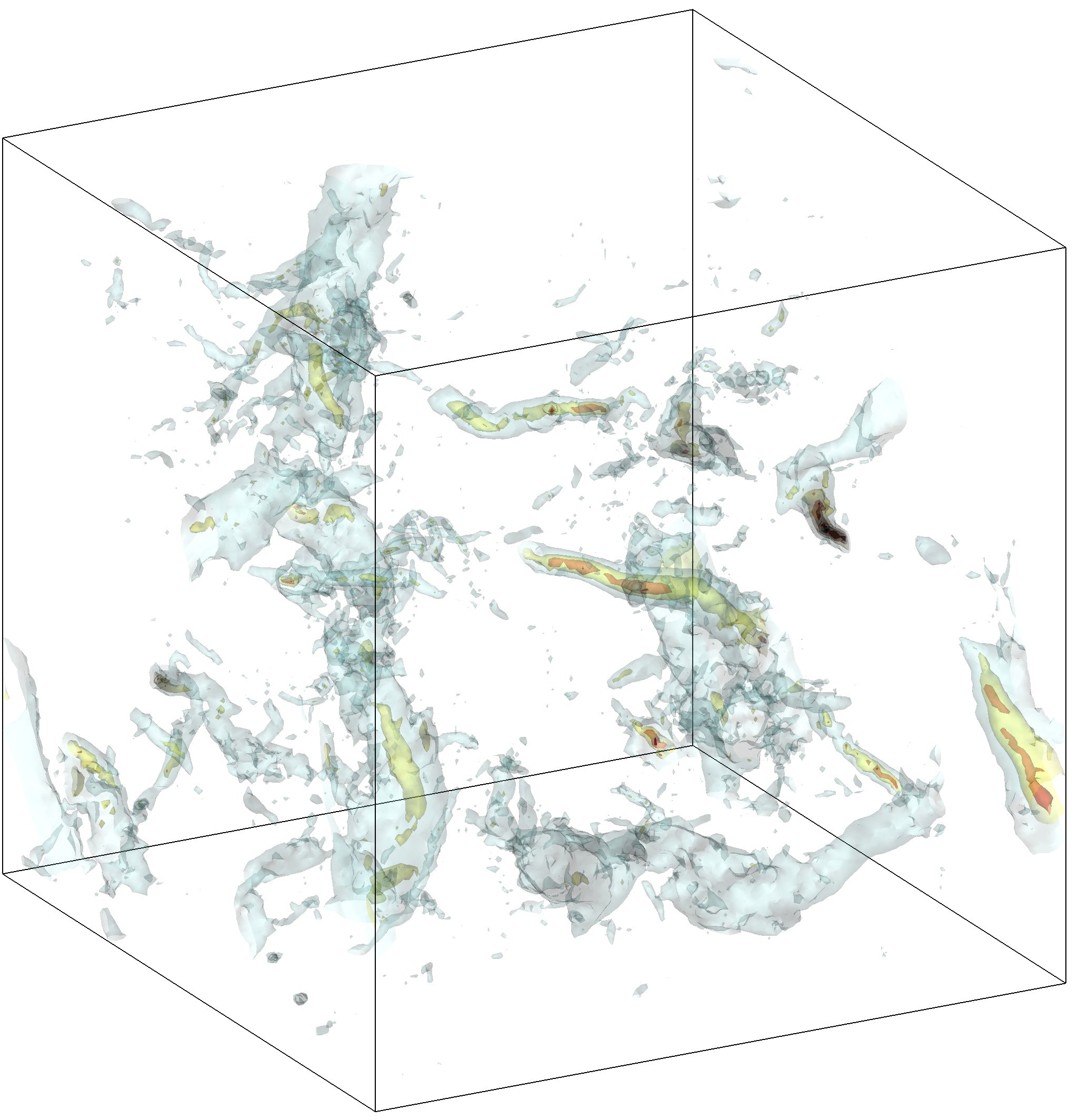}
\caption{Pressure isosurfaces of value $p=-2$, -4, -6, -8 and -10 during a burst of very-low-pressure events. The corresponding times are, from left to right and top to bottom, $t=3.8$, 6.4, 9.1, 11.8, 14.4 and 25.1. The size of the periodic box is 55.6 and the integral scale is 12.1.}\label{fig:burststructure}
\end{figure*}

        The high burstiness of very-low-pressure events is an indication of large flow structures being involved in them, such that when one of these structures appears many particles go through it and bursts of events take place. This picture is consistent with the intermittent structures of intense vorticity, or {\em worms}, first described by Jim\'enez {\em et al} \cite{jwsr93_jfm}, which have lengths of the order of the integral scale of the flow. To confirm this, we looked at the pressure isosurfaces of the $\Re_\lambda=418$ simulation, for which a burst of low-pressure events takes place for non-dimensional times between 6 and 18 (the burst thus lasts about 46 Kolmogorov timescales). The isosurfaces are shown in Fig. \ref{fig:burststructure} at some selected instants. The lightest surfaces, corresponding to $p=-2$, are present in all snapshots as expected from the low value of $B$ for $p_-=-2$. At time $t=6.4$ a vortical structure develops, which is most evident at the peak of the burst (between $t=9.1$ and 11.8, third and fourth frames in the figure). It is within this structure that very low pressures (-8 or lower) occur and affect numerous Lagrangian particles. By time 14.4 this structure is dissolving away and after the burst, at $t=25.1$, the flow has recovered an isosurface pattern similar to the one observed before the burst. We have checked that the vertical low-pressure structure is indeed a high vorticity region, and its shape and length are in agreement with the intermittent worms reported in the literature (notice that the integral lengthscale for this flow is 12.1, roughly 1/5 the box edgelength) \cite{jwsr93_jfm}.

\section{Conclusions}

Motivated by cavitation inception modeling, this work reports some Lagrangian statistics of the pressure field in forced homogeneous isotropic turbulence. It is clear that for a cavitation nucleus to grow up to detectable size a pressure fluctuation that takes it to sufficiently low pressures for long enough time is necessary. However, the frequency of such low-pressure events was not available in the literature and is first reported herein for two values of $\Re_\lambda$, namely 150 and 418. The main result consists of the average frequency $\zeta(p_-,d)$ with which a Lagrangian particle undergoes a fluctuation that takes { its pressure} below some threshold $p_-$ for a time longer than some minimum duration $d$. This average frequency, for any $d$, is observed to have an exponential tail (i.e., to behave as $\sim C \exp(\beta p_-)$) towards very low pressure thresholds. Furthermore, the value of the logarithmic slope $\beta$ that corresponds to $d\simeq 0$ is roughly coincident with that of the exponential tail of the pressure PDF. The PDF of the duration of the low-pressure events is also reported, which shows that the most probable duration is smaller than the Kolmogorov timescale and quite insensitive to $p_-$ for both $\Re_\lambda$ considered. On the other hand, the mean and median duration of the pressure excursions grow significantly with $\Re_\lambda$ and depend strongly on $p_-$, but only for moderate values of this variable. The analysis of the interarrival times between low-pressure events shows that their occurrence departs from that of a totally random homogeneous stochastic process (Poisson process). This departure becomes more and more accentuated as the threshold $p_-$ is lowered. The distribution of interarrival times is heavy-tailed, indicative of a bursty process. In fact, a quantitative indicator of burstiness was computed, yielding values indicative of a highly intermittent process. This suggests that the bursts of low-pressure events are associated with intermittent large-scale vortical structures\cite{jwsr93_jfm}, as confirmed by examination of the pressure isosurfaces at the time of the bursts.

The reported results provide useful quantitative data to predict the frequency, intensity and duration of pressure fluctuations experienced by very small particles that are passively transported by a turbulent flow. {They can be used, for example, to inform modern models of numerical cavitation \cite{co18_cav,lha18_cav,mc18_jcp}}. The behavior at higher Reynolds numbers and in other turbulent flows should certainly be explored to gain further understanding. Also, the relative velocity that develops between non-neutrally-buoyant particles of finite size and the surrounding liquid could attract bubbles toward vortex cores and strongly affect the computed frequencies. These issues are the subject of ongoing work.

\bigskip

\section*{Acknowledgments}

The authors are thankful to Adri\'an Lozano-Dur\'an for his help in accessing and processing the database at Univ. Polit\'ecnica de Madrid. This research was sponsored by the US Office of Naval Research through MURI grant N00014-17-2676, Univ. of Minnesota lead institution, Dr. Ki-Han Kim program manager.
GCB acknowledges support from the S\~ao Paulo Research Foundation (FAPESP, Brazil), grant 2018/08752-5.

\end{document}